\documentclass[preprint]{revtex4-2}
\usepackage[english]{babel}
\usepackage{amsmath}
\usepackage{amssymb}
\usepackage{bm}

\usepackage{color}
\usepackage{epsfig}

\newcommand{\nix}[1]{}
\newcommand{\addELI}[1]{\textcolor{red}{#1}}

\usepackage{mathrsfs}
\usepackage{ulem}
\usepackage[scr=rsfs,cal=boondox]{mathalfa}

\begin{document}

\title{	
		Electron and hole $g$ factors 
		in semiconductors \\ and nanostructures (Review)
	}
	\author{A.V. Rodina, M.A. Semina, E.L. Ivchenko}
	\affiliation{Ioffe Institute, St. Petersburg, Russia}

\begin{abstract}
	We present a review of experimental and theoretical studies of the spin response of charge carriers to an external magnetic field in bulk semiconductors and semiconductor nanostructures. The linear response is quantitatively characterized by the magnitude of the electron or hole $g$ factor. Various experimental methods for measuring the electron $g$ factor are considered, beginning with historical works and including modern research. A detailed analysis of theoretical methods for calculating the electron and hole $g$ factors in bulk semiconductors and nanostructures of various shapes also includes fundamental work from previous years and the present time.
	
\end{abstract}
\date{November 4, 2025}
\maketitle

	\section{Introduction}
The year 2025 marks the 100th anniversary of the discovery of electron spin \cite{Spin}. One of the most important characteristics of electron spin is the $g$ factor. This review examines the $g$ factor of charge carriers in semiconductors and semiconductor nanostructures.

In works on measuring the $g$ factor of a free electron \cite{Kusch,Crane1954, CraneReview,Ekstrom} it is defined as a positive coefficient
\begin{equation} \label{g0}
	g_0 = 2 (1 + a) \approx 2.002319\dots\:,
\end{equation}
where, taking into account quantum electrodynamic corrections, the coefficient $a = \alpha/(2 \pi) + \dots$, $\alpha$ is the fine structure constant. The same sign of $g_0$ for a free electron in a vacuum is chosen in solid-state physics when studying the change in the Land\'{e} factor, or $g$ factor, taking into account its renormalization by the spin-orbit interaction. In this case, the magnetic and mechanical moments, ${\bm m}$ and ${\bm s}$, respectively, are related by 
\begin{equation} \label{mg0}
	{\bm m} = - g_0 \mu_B {\bm s}\:,
\end{equation} 
where $\mu_B$ is the  Bohr magneton (positive value). Note that in physics, $g_0$ is sometimes considered to be a negative value; see, for example, Ref.~\cite{sign}. In this case, the minus sign is missing from Eq.~(\ref{mg0}).

In this review, we adhere to the definitions (\ref{g0}) and (\ref{mg0}). Then the effect of the magnetic field ${\bm B}$ on the electron spin in a doubly spin-degenerate conduction band, or the Zeeman effect, is described by the Hamiltonian
\begin{equation} \label{HZ}
	{\cal H}_Z = \frac12 g_{ij} \mu_B \sigma_i B_j\:,
\end{equation} 
where $i,j$ are the Cartesian coordinates $x,y,z$, $\sigma_i$ are the Pauli spin matrices, $g_{ij}$  is the tensor of $g$ factors. In bulk cubic crystals, as well as in spherical quantum dots grown from composite materials with the cubic symmetry, there is only one linearly independent component of the  $g$ factor tensor:
\begin{equation} \label{g} 
	g_{ij} = g \delta_{ij}, 
\end{equation} 
and the operator (\ref{HZ}) takes the form $g \mu_B {\bm \sigma} \cdot {\bm B}/2$. In symmetric structures with a GaAs/GaAlAs quantum well grown along the $z\parallel[001]$ axis (point group D$_{2d}$), there are two linearly independent components $g_{zz} \equiv g_{\parallel}$ and
$g_{xx} = g_{yy} \equiv g_{\perp}$. As the symmetry of the nanostructure decreases, the number of linearly independent components increases. In a quantum dot of complex shape, all 9 components $g_{ij}$ can be different.

In the next section, we present a variety of methods for measuring the $g$ factor. The second part of the review focuses on the theory of the electron Zeeman effect in bulk crystals (Section 3), two-dimensional heterostructures (Section 4), and quantum dot nanostructures (Section 5). Section 6 examines the hole Zeeman effect. The final Section 7 summarizes the results and outlines future prospects.

\section{Methods for measuring the $g$ factor} 
\subsection{Electron spin, or paramagnetic, resonance (ESR or EPR)} 
Electron spin resonance was first discovered by E.K. Zavoisky in 1944 on the crystal hydrate MnSO$_4$$\cdot$7H$_2$O at a frequency of an alternating magnetic field $\nu \approx 10$ MHz \cite{Zavoiskii}. He proceeded from the fact that the absorption of a high-frequency field is proportional to the imaginary part of the magnetic susceptibility \cite{Zavoiskii}
\begin{equation} \label{chi} 
	\chi'' = \frac{2 \nu_0^2 \nu \nu' \chi_0}{ (\nu_0^2 - \nu^2)^2 + 4 \nu^2 \nu^{\prime 2}} \:,
\end{equation}
where $\chi_0$ is the static magnetic susceptibility, $\nu_0$  is the Larmor precession frequency, $\nu'$  is the damping. The formula (\ref{chi}) describes the resonant response at the frequency determined by the relation
\begin{equation} \label{gmB} 
	h \nu_0  = \hbar \omega_0 = |g| \mu_B B \:,
\end{equation}
where $B$  is the magnetic field, and the coefficient $g$ is the $g$ factor introduced in Section 1. Thus, measuring the resonant frequency $\nu_0$ or $\omega_0$ allows one to unambiguously determine the modulus of the $g$ factor. Note that the EPR (or ESR) spectrum is the dependence of the measured signal not only on the frequency of the alternating field, but also on the magnitude of the magnetic field causing the splitting of the spin sublevels. Moreover, in the case of continuous wave excitation, it is often not the resonant absorption line that is recorded, but the derivative of this line. This allows for a more accurate determination of the resonant magnetic field value corresponding to the intersection of the first derivative with the zero line, as well as the determination of the line width by the distance between the maximum and minimum points.

The experimental study of EPR in semiconductors was initiated by J. Bemski. Using microwave radiation at frequencies of 9 and 24 GHz, he discovered a feature in the dependence of absorption on the external magnetic field, determined by the resonant spin flip of free electrons at the Fermi surface in $n$-type InSb samples with electron concentrations from 2$\times$10$^{14} to 3\times$10$^{15}$ cm$^{-3}$. The determined modulus of the electron $g$ factor varied from 50.7 to 48.8. Isaacson R.A.  extended the concentration range to $3.6 \times 10^{13}$ -- $1.5 \times 10^{15}$ cm$^{-3}$ and measured the dependence of the parameter $|g|$ on the Fermi energy of electrons \cite{InSb}, which was in satisfactory agreement with the formula derived in the ${\bm k}\cdot {\bm p}$ perturbation theory for semiconductors with zinc blende structure \cite{Roth}, see details in Section \ref{bulk}.

\subsection{Optically detected magnetic resonance (ODMR)}
\begin{figure}[h!]
	\centering
	\includegraphics[scale=0.5]{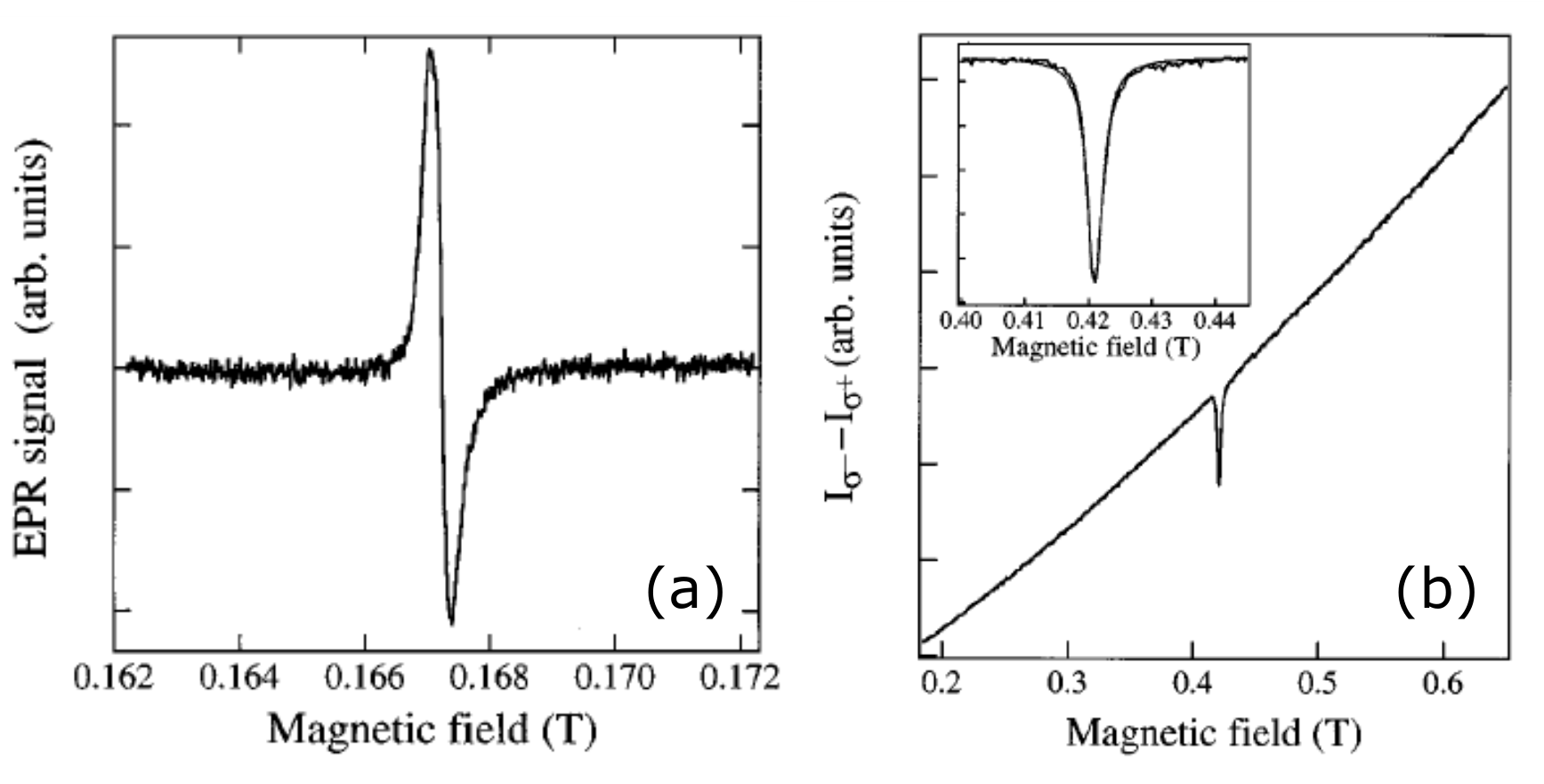}
	\caption{ (a) EPR spectrum ($f_{mw}$ = 9.5 GHz, $T=4.5$ K) of a bulk In$_{0.53}$Ga$_{0.47}$As crystal, recorded from the first derivative of the resonant absorption line. (b) ODMR spectrum ($f_{mw}$ = 24 GHz, $T =1.6$ K), recorded from the magnetically induced circular polarization of the exciton PL of a bulk In$_{0.53}$Ga$_{0.47}$As crystal. The inset shows the resonance after subtracting the linear background and approximating it using a Lorentzian contour. Figure adapted from Ref.~\cite{Kowalski}.}
	\label{Fig1} 
\end{figure}
The optically detected magnetic resonance (ODMR) method is a combination of EPR and photoluminescence (PL). This method records changes in the PL intensity and polarization as a function of the microwave frequency and/or an applied external magnetic field. This allows determination of the frequencies of resonant transitions between spin sublevels of a charge carrier or exciton. Figure \ref{Fig1} shows the EPR (a) and ODMR (b) spectra measured in the bulk semiconductor $n$-In$_{0.53}$Ga$_{0.47}$As solid solution ($n$ = $2 \times 10^{15}$ cm$^{-3}$) \cite{Kowalski}. Spectrum (a) was measured at a frequency $\nu = 9.5$ GHz at a temperature $T = 4.5$ K with the first derivative recorded. Figure \ref{Fig1}(b) presents the dependence of the intensity difference $I_{\sigma^-} - I_{\sigma^+}$ of the circularly polarized components of exciton PL on the magnetic field. To record the spectrum (b), a microwave field frequency $\nu = 24$ GHz was used, exciton luminescence was excited nonresonantly by laser radiation at a wavelength of 514 nm. The obtained values of (a) $|g| = 4.0746 \pm 0.005$, related to conduction band electrons or electrons bound to a shallow donor, and (b) $|g| = 4.07 \pm 0.02$, related to electrons bound into an exciton, are practically identical. This indicates that the $g$ factor does not change when a conduction band electron is bound into an exciton.

Unlike EPR, optical detection of the electron resonance  does not require a large number of spins in the sample. Reliable measurement of the luminescence signal is sufficient, provided that its intensity or polarization depends on the spin polarization of electrons or excitons, which is changed under the effect of a microwave field. In principle, it is possible to observe ODMR on a single exciton. Therefore, this method is in demand for studying the Zeeman effect in semiconductor nanostructures, for example, in GaAs/AlAs type I and II superlattices \cite{Kesteren,Romanov1994,Baranov1994}.

\subsection{Spin-flip Raman scattering}
During inelastic spin-flip scattering of light (spin-flip Raman scattering) in an external magnetic field, the initial and final states of the system differ in the spin states of an electron, hole, or exciton. As a result, the energies of the incident ($\hbar \omega$) and scattered ($\hbar \omega'$) photons differ by the Zeeman splitting energy $\Delta_Z = |g| \mu_B B$: for scattering into the Stokes (low-frequency) and anti-Stokes (high-frequency) regions, the so-called Raman shift is given by $\hbar (\omega - \omega') = \pm \Delta_Z$, respectively. This phenomenon was predicted by Yafet \cite{Yafet} and first observed for free electrons in $n$-InSb \cite{SFexp} and for bound electrons and holes in CdSe \cite{Hopfield1968}. In addition to single spin-flip scattering, it is possible to observe double and triple scattering processes, in which the spin direction changes for two or three electrons \cite{Damen1972,Economou1972,Cardona1981,Kudlasik2020} and the Raman shift is $\pm 2 \Delta_Z$ or $\pm 3 \Delta_Z$.

\begin{figure}[h!]
	\centering
	\includegraphics[scale=0.5]{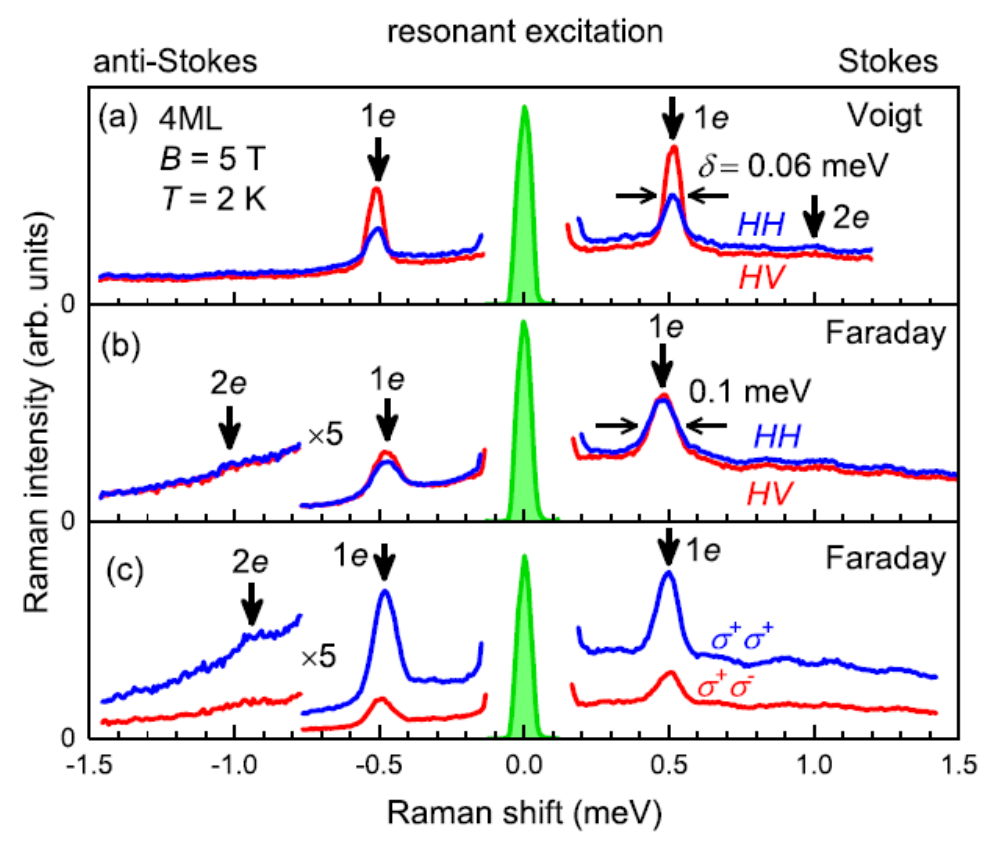}
	\caption{(a -- c) Spin-flip Raman spectra of 4 monolayer thick CdSe nanoplatelets measured under resonant excitation at $\hbar \omega =2.497$ eV, power density P = 20 W cm$^{-2}$, $B$ = 5 T, and $T$ = 2 K. (a) Spectra in  Voigt geometry  measured in co- (blue) and cross- (red) linear polarizations. Faraday spectra measured in co- and cross-linear polarizations [panel (b)] and in co- and cross-circular polarizations [panel (c)]. Figure adapted from Ref. \cite{Kudlasik2020}.}
	\label{Fig2} 
\end{figure}

In nanostructures, double electron spin flip scattering   was, for the first time, observed in CdSe colloidal nanoplatelets (two-dimensional nanoplates) in Ref.~\cite{Kudlasik2020}. Spin-flip Raman scattering spectra (SFRS) for CdSe nanoplatelets 4 monolayers thick are shown in Fig.~\ref{Fig2}. The theory of single and double electron spin flip Raman scattering   in semiconductor nanoplatelets is developed in Ref.~\cite{RI2020}. Colloidal CdSe nanoplatelets, the thickness of which is several monomolecular layers, and the lateral dimensions are several tens of nanometers, can stand vertically on a silicon substrate, lie horizontally on it, or be tilted at some angle \cite{Kudlasik2020}. In \cite{RI2020} a microscopic theory is constructed and the following selection rules are derived for Raman scattering involving one ($1e$) and two ($2e$) electrons
\begin{subequations}
	\begin{align}
		\label{eq_syst1}
		&I^{(1e)} \propto \sin^2{\tilde{\theta}} \left\vert \left( {\bm e}^* \times {\bm e}^0 \right)\cdot {\bm c} \right\vert^2\:, \\ & \label{eq_syst2}
		I^{(2e)} \propto \sin^4{\tilde{\theta}} \left\vert {\bm e}^* \cdot {\bm e}^0 - ({\bm e}^* \cdot {\bm c}) ({\bm e}^0 \cdot {\bm c}) \right\vert^2 \:.\end{align}
\end{subequations}
Here $I^{(ne)}~(n=1,2) $  is the intensity of the secondary radiation, ${\bm e}^0$ and ${\bm e}$ are the unit polarization vectors of the primary and secondary beams (${\bm e}^*$  is the complex conjugate of ${\bm e}$), ${\bm c}$  is the unit vector in the direction of the normal to the platelet, $\tilde{\theta}$  is the angle between the vector ${\bm c}$ and the direction of the effective magnetic field in the nanoplatelet $\left( g_{\parallel} {\bm B}_{\parallel} + g_{\perp} {\bm B}_{\perp}\right)/g$, where $g_{\parallel}$ and $g_{\perp}$ are the electron $g$ factors for the longitudinal (${\bm B} \parallel {\bm c}$) and transverse (${\bm B} \perp {\bm c}$) directions of the external magnetic field ${\bm B}$. Here it is taken into account that in the presence of anisotropy, $g_{\parallel} \neq g_{\perp}$, the magnitude of the Zeeman splitting is $g \mu_B B$, where
\begin{equation}  \label{geff}
	g = \sqrt{g_\perp^2 \sin^2{\theta_B} +g_\parallel^2  \cos^2{\theta_B} }\: 
\end{equation} 
and $\theta_B$ is the angle between the vectors ${\bm c}$ and ${\bm B}$.
The angles $\tilde{\theta}$ and $\theta_B$ are related by 
$$
\sin^2{\tilde \theta} = \frac{g_\perp^2}{g^{2} } \sin^2{\theta_B} \, .
$$

The scattering mechanism considered in deriving Eqs.~(\ref{eq_syst1}), (\ref{eq_syst2}) includes the absorption of a primary photon with the excitation of an exciton in the platelet, the spin flip of one or two  localized resident  electrons due to the exchange interaction with the electron in the exciton, and the emission of a secondary photon by the exciton. From a comparison of the selection rules with experiment, it was concluded that the nanoplatelets mainly lie on the substrate or are located at a small angle to it. In the further experimental studies \cite{Meliakov2023} the dependence of the electron $g$ factor on the angle between the magnetic field direction and the $\bm c$ axis in CdSe nanoplatelets was measured. This dependence is well described by the expression (\ref{geff}), and thus the  anisotropy of the  $g$ factor was determined.

Spin-flip  Raman scattering  in semiconductor organo-inorganic and inorganic perovskites was experimentally and theoretically studied in Refs.~\cite{NatCom2022,RI2022,RI2024}. In the lead halide perovskite family, a universal dependence of the electron and hole $g$ factors on the band gap was established, while their anisotropy was determined in perovskites with symmetry below cubic. The mechanism and feasibility of observing exciton spin-flip scattering of light by an acoustic phonon in cubic perovskites are discussed in \cite{RI2024}. The Raman shift is determined both by the total value and half-value of exciton $g$ factors.

\subsection{Spectral resolution of exciton sublevels} 

The most direct way to determine the exciton $g$ factor is to observe the splitting of reflection, absorption, or luminescence spectral lines in a magnetic field. One such possibility arises in dilute magnetic semiconductors, or semiconductor solid solutions, in which some of the atoms in one of the sublattices are replaced by paramagnetic impurity atoms, such as Mn, Fe, or Co. An external magnetic field induces a magnetic moment in the paramagnetic ions. The $sp$-$d$ exchange interaction of magnetic ions with electrons and holes leads to a giant spin splitting of the charge carrier states, which by far exceeds the Zeeman splitting in the original nonmagnetic matrix. As a result, several separate lines appear in the optical spectra, the amplitude of which is polarization-dependent \cite{Ryabchenko}.

\begin{figure}[h!]
	\centering
	\includegraphics[scale=0.5]{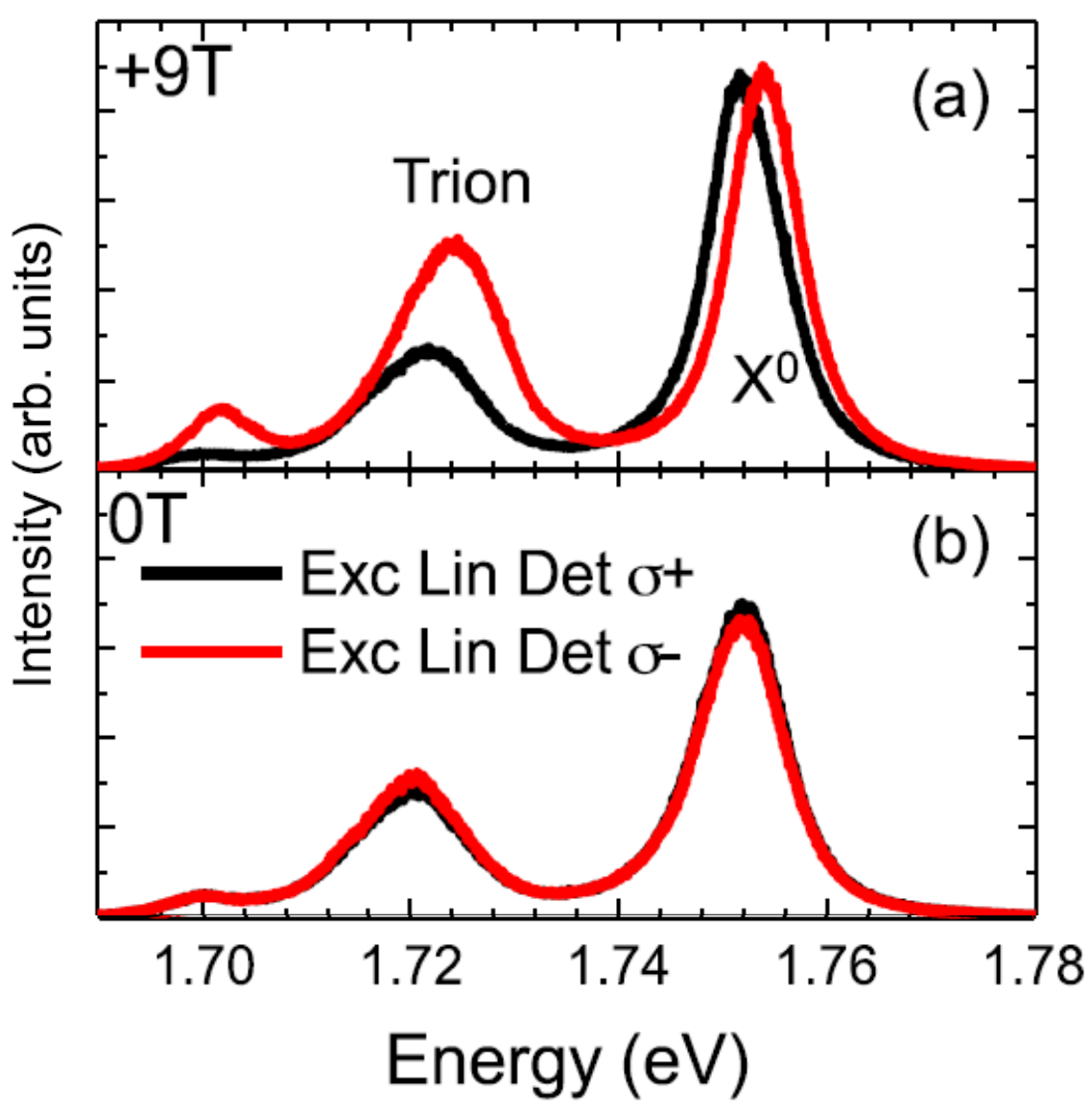}
	\caption{Photoluminescence spectra measured in $\sigma_+$ (black) and $\sigma_-$ (red) circular polarization in a WSe$_2$ monolayer under linearly polarized photoexcitation. Panel (b) shows experimental curves in the absence of a magnetic field. The symbols $X^0$ and ``Trion'' denote the emission lines of the neutral exciton and the negatively charged exciton X$^{-}$ (trion). The monolayer lies on a SiO$_2$/Si substrate. Figure adapted from Ref. \cite{TMDC}.} 
	\label{Fig3} 
\end{figure}

Another possibility can be realized in the case of narrow exciton lines whose half-width is smaller than or comparable to the Zeeman splitting of the exciton sublevels. This possibility is illustrated in Fig. \ref{Fig3}, which shows the magnetic-field splitting of the exciton PL spectrum (peak $X^0$) in a monomolecular WS$_2$ layer \cite{TMDC}. When the magnetic field direction is inverted, $B_z \to - B_z$ the spectral positions of the circularly polarized luminescence peaks $\sigma_{\pm}$ swap.

\subsection{Magnetic circular polarization of luminescence (MCPL)}

If the spin relaxation time between the Zeeman sublevels of an exciton is shorter than or comparable to the lifetime of the exciton or exciton complex, the spin sublevels in an external magnetic field will be selectively populated, and the PL will be circularly polarized, at least partially, even under unpolarized or linearly polarized excitation conditions; see review \cite{MCPL} for more details. This possibility is realized in the emission spectrum of a trion in a WSe$_2$ monolayer (Fig. \ref{Fig3}, ``Trion'' peak). As can be seen from the figure, the exciton PL peaks coincide in intensity since the lifetime $\tau_0$ of an exciton $X_0$ is short compared to its spin relaxation time $\tau_s$. For the trion, the inverse relation $\tau_s < \tau_0$ holds, and the intensity of one of the circularly polarized PL components significantly exceeds the other.

The advantage of the MCPL method is its relative simplicity. In particular, it does not require spectral resolution of the exciton Zeeman sublevels and can be applied to a wide class of objects with broad spectral lines or PL bands. In Ref.~\cite{Kotova}, scenarios for the MCPL formation in an inhomogeneous ensemble of localized excitons are theoretically studied in the case where the exciton $g$ factor strongly depends on its energy. The MCPL method is one of the main methods for determining the  $g$ factors of excitons and trions in ensembles of colloidal nanostructures in which the inhomogeneous PL line width does not allow direct determination of the splitting of Zeeman sublevels \cite{JohnstonHalperin2001,Liu2013,Shornikova2020,Qiang2021}. At low temperatures, it is possible to determine the $g$  factor of a dark (spin-forbidden, optically inactive) exciton, as well as the $g$ factor of a heavy hole in a negatively charged trion. In the work \cite{Shornikova2020}, using the MCPL method for CdSe nanoplatelets, it was shown that the exchange interaction of an exciton with surface-bound charge carriers or dangling bonds significantly contributes to its Zeeman splitting and controls not only the magnitude but also the sign of the MCPL.

\subsection{Hanle effect} 

Under the interband absorption of circularly polarized light, the projection of the angular momentum of photons is transferred to the excited electrons and holes. As a result, the total average spin ${\bm s}_0$ of photoelectrons at the moment of birth is nonzero. During its lifetime in the conduction band, this spin experiences Larmor precession in a transverse magnetic field ${\bm B}$ with the angular frequency ${\bm \Omega}_L = g \mu_B {\bm B}/\hbar$. The spin dynamics of the  optically spin-oriented electrons is described by the kinetic equation
\begin{equation} \label{Bloch}
	\frac{d {\bm s}}{d t} + \frac{\bm s}{T} + {\bm s} \times {\bm \Omega}_L = {\bm G}_s\:,
\end{equation}
where $T$  is the spin lifetime determined by the lifetime and spin relaxation time of the photoelectron:
\begin{equation} \label{T}
	\frac{1}{T} = \frac{1}{\tau_0} + \frac{1}{\tau_s} \:,
\end{equation}
${\bm G}_s$ is the  generation rate of electron spins. In what follows, we will assume that ${\bm G}_s \parallel z$ and ${\bm B} \perp z$. For simplicity, we neglect in Ref.~(\ref{Bloch}) the effect of the interaction of the electron spin with the fluctuating spin of the main lattice nuclei.

Under steady-state excitation conditions, when ${\bm G}_s \parallel z$ is time-independent and $d{\bm s}/d t = 0$, the equation for the average spin ${\bm S}= {\bm s}/N$ of one photoexcited electron ($N$ is the steady-state number of photoexcited electrons) coincides with Eq.~\eqref{Bloch} for ${\bm s}$, in which ${\bm G}_s$ should be replaced by ${\bm G}_s/N$. In this case, we obtain for the projection of ${\bm S}$ onto the $z$ axis
\[
{S}_z (B_{\perp}) = \frac{{S}_z(0)}{1 + (\Omega_L T)^2}\:, 
\]
where ${ S}_z(0)$ is the projection onto the $z$ axis of the average spin per electron generated under constant  (CW) circularly polarized excitation in zero magnetic field:
\begin{equation} \label{S0}   { S}_z(0) = \frac{{ s}_z(0)}{N}  = \frac{T}{\tau_0}\frac{G_{sz}}{N} = \frac{\tau_s}{\tau_s+\tau_0}\frac{G_{sz}}{N} \, .
\end{equation}
Taking into account the relationship between the selection rules during generation and radiative recombination, the degree of circular polarization of PL is described by the Lorentzian
\begin{equation} \label{circ}
	P_{\rm circ}(B) = \frac{I_{\sigma_+} - I_{\sigma_-}}{I_{\sigma_+} + I_{\sigma_-}} = \frac{P_{\rm circ} (0)}{1 + (B/B_{1/2})^2}\:,
\end{equation} 
where $B_{1/2} = \hbar/(|g| \mu_B T)$. While deriving expression (\ref{circ}), it was assumed that spin relaxation of the photoholes is very fast and they are unpolarized. Depolarization of PL in a transverse magnetic field is called the Hanle effect. Measuring the dependence $P_{\rm circ}(B) $ allows one to find the product $|g|T$. In this case, the condition for observing the Hanle effect is opposite to the condition for observing the MCPL: the spin relaxation time of the photoexcited electron $\tau_s$ should not be too small compared to its lifetime $\tau_0$. Note that, strictly speaking, the spin relaxation time $\tau_s$, which appears in the ratio $T/\tau_0$ in the expression \eqref{S0} for the average spin in zero magnetic field, and the time $\tau_s$, which determines the effective magnetic field $B_{1/2}$, may differ. They correspond to the longitudinal and transverse (or spin coherence) spin relaxation times, respectively. In an ensemble of quasi-particles, the spin coherence time is determined by the spin dephasing caused by the spread of parameters. Figure 1 in \cite{Dzioev} and Fig. 2 in \cite{Kalevich} can serve as examples of Hanle effect observations.

\subsection{Spin beats}

To describe the spin dynamics of electrons excited by a short circularly polarized pulse, we must set ${\bm G}_s = 0$ in Eq.~(\ref{Bloch}), but take into account the time derivative $d {\bm s}/{d t} \ne 0$ and introduce initial conditions for the total average spin ${\bm s}_0$ and the number of photoexcited electrons $N_0$. For the spin component along the $z$ axis, beats appear in a transverse magnetic field at the Larmor precession frequency: 
\begin{equation} \label{St}
	s_z(t) = {\rm e}^{- t / T_2} \cos{(\Omega_L t)} s_{0,z}\, , \quad  S_z(t) = {\rm e}^{- t /\tau_s} \cos{(\Omega_L t)} S_{0,z}\,  ,
\end{equation} 
where $S_{0,z} = s_{0,z}/N_0$, and the time $T_2$ is determined by the lifetime $\tau$ and the spin coherence time $\tau_s$.

\begin{figure}[h!]
	\centering
	\includegraphics[scale=0.18]{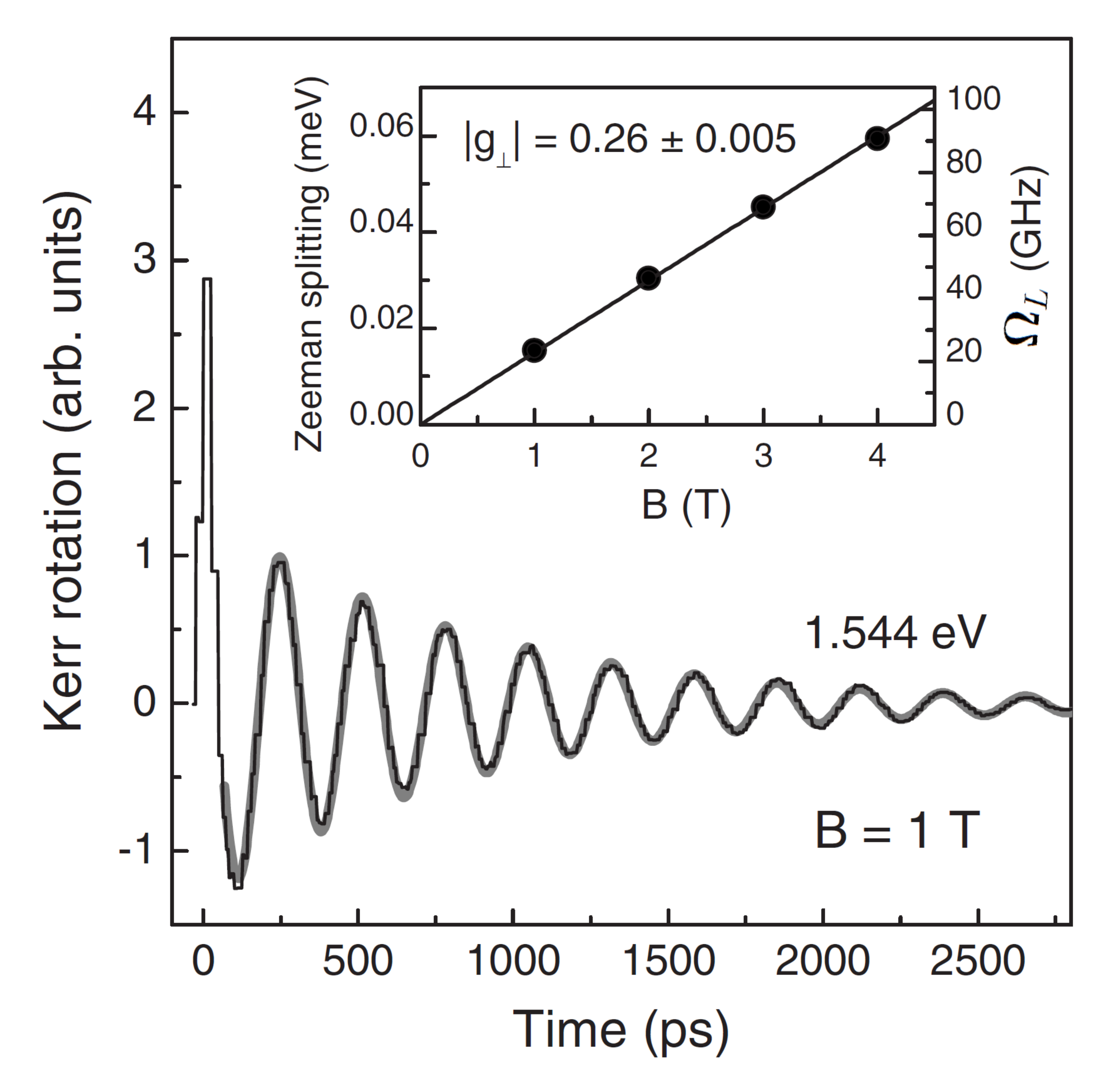}
	\caption{Kerr rotation as a function of time in a 10 nm thick GaAs/Al$_x$Ga$_{1-x}$As quantum well in a magnetic field of 1 T and at T=1.6 K. The black line shows the experimental data, the thick gray line corresponds to the results of the approximation according to Eq.~(\ref{St}) with parameters $\Omega_L=23.4$ GHz and $T_2=\tau_s$ =880 ps. The inset shows the Zeeman splitting (left scale) and the spin beat frequency $\Omega_L$ (right scale) as functions of the magnetic field.
Figure adapted from Ref.~\cite{Kiselev2007}. }
	\label{Fig4} 
\end{figure}

Currently, a very sensitive and efficient pump-probe method is used to detect spin beats. An incident  pump pulse is circularly polarized and is followed by a pulse of linearly polarized probe light, delayed by $t$. Due to the spin Kerr effect, the plane of polarization of the  probe pulse  reflected from the sample is rotated by an angle $\delta \theta \propto s_z(t)$, the magnitude of which is measured by a polarization-sensitive balanced detector. More details on this method can be found in the review \cite{GlazovReview} and in Chapter 15 of the collective monograph \cite{Manfred}.

Figure \ref{Fig4} illustrates the electron spin beats observed in a GaAs/Al$_x$Ga$_{1-x}$As quantum well \cite{Kiselev2007}. They allow one to determine with high accuracy the value of the transverse electron $g$ factor $|g_{\perp}| = 0.26 \pm 0.005$. In this case, the lifetime of resident electrons in the quantum well is assumed to be infinite, and the spin lifetime $T_2$ is determined by the spin coherence time. The values of the longitudinal electron $g$ factor $|g_{\parallel}|$ in a quantum well can be determined by recording spin beats in a magnetic field inclined to the well growth axis \cite{Kiselev2007}. Note that in an ensemble of randomly oriented two-dimensional colloidal nanoplatelets, the measured Kerr rotation signal is always determined by the magnitude of the transverse electron $g$ factor, and the presence of its anisotropy makes only a small contribution to spin dephasing, accelerating the attenuation of the signal amplitude \cite{Meliakov2023,Golovatenko2025}. Moreover, in an ensemble of randomly oriented nanocrystals, it is also possible to observe spin beats in a magnetic field parallel to the direction of the light beam propagation (Faraday geometry) \cite{Gupta2002}. 

\subsection{Anticrossing of exciton sublevels}

We illustrate the exciton sublevel anticrossing method using the example of an exciton $X^0_{e\mbox{-}hh}$ formed from an electron in the conduction band with spin $s = \pm 1/2$ and a heavy hole with internal angular momentum projection $j = \pm 3/2$ \cite{MCPL}. Of the four exciton states $|s, j\rangle=|\pm 1/2, \pm 3/2\rangle$, two states with a total angular momentum projection $m = s + j = \pm 1$ are optically active (light), while states with $m = \pm 2$ are inactive (dark). In structures with reduced C$_{2v}$ symmetry, the dependencies of the exciton sublevel energies on the magnetic field ${\bm B} \parallel z$ are described by the expressions \cite{MCPL,Kaminskii}
\begin{eqnarray}
	&&E_{1,2} = E_0 + \frac12 \left( \delta_0 \pm \sqrt{\delta_2^2 + (g_e - g_{hh})^2\mu_B^2 B_z^2}\right)\:,\\
	&&E_{3,4} = E_0 + \frac12 \left( - \delta_0 \pm \sqrt{\delta_1^2 +  (g_e + g_{hh})^2 \mu_B^2 B_z^2} \right)\:. \nonumber
\end{eqnarray}
Here $E_0$ is the exciton excitation energy without taking into account the exchange interaction and at $B_z = 0$, $g_e$ and $g_{hh}$ are the longitudinal $g$ factors of the electron and hole, $\delta_0, \delta_1$ and $\delta_2$ are the exchange splitting constants, usually $\delta_0 \gg \delta_1, \delta_2$. Pairs of sublevels 1, 2 and 3, 4 represent bright and dark exciton states split by the anisotropic exchange interaction and the magnetic field. In this case, the lifetime of the bright exciton states is shorter, since both radiative and nonradiative processes of exciton recombination contribute to it. In a magnetic field, excitons 1 and 2 emit elliptically polarized photons with opposite signs of circular polarization.

If the condition $g_e + g_{hh} > g_e - g_{hh}$ is fulfilled, sublevel 3 intersects sublevels 1 and 2 with increasing field $B_z$ at some values of $B_{{\rm cr},1}$ and $B_{{\rm cr},2}$. Any additional small local reduction in symmetry leads to the mixing of states 3 and 1 (or 2) in the vicinity of the critical field point $B_{{\rm cr},1}$ ($B_{{\rm cr},2}$), so that both mixed states become bright and their lifetime exceeds the lifetime of the bright state 2 (or 1). As a result, the exciton PL intensity increases and it becomes partially circularly polarized. This effect was observed experimentally, in particular, on GaAs/AlAs superlattices in the works \cite{Kesteren, Romanov1994, Baranov1994}, and  the theory was constructed in Ref.~\cite{Kaminskii}. 

\subsection{Electron spin fluctuations (spin noise)}

Even in the absence of any average spin orientation in a sample with $N$ electrons, a fluctuation spin $s_z$ of the order of $\sqrt{N}$ is expected to exist due to the randomly directed spins of individual electrons. Therefore, similarly to the situation in the pump-probe method, the probe linearly polarized light experiences a rotation of the plane of polarization by an angle $\theta \propto s_z(t)$ when passing through the sample without any pumping. In a transverse magnetic field, a peak in the spin noise spectrum appears  at the Larmor precession frequency $\Omega_L$, determined by the electron $g$ factor, see e.g. \cite{GlazovIvch2012,Oestreich2014,Smirnov2014}.

\begin{figure}[h!]
	\centering
	\includegraphics[scale=0.51]{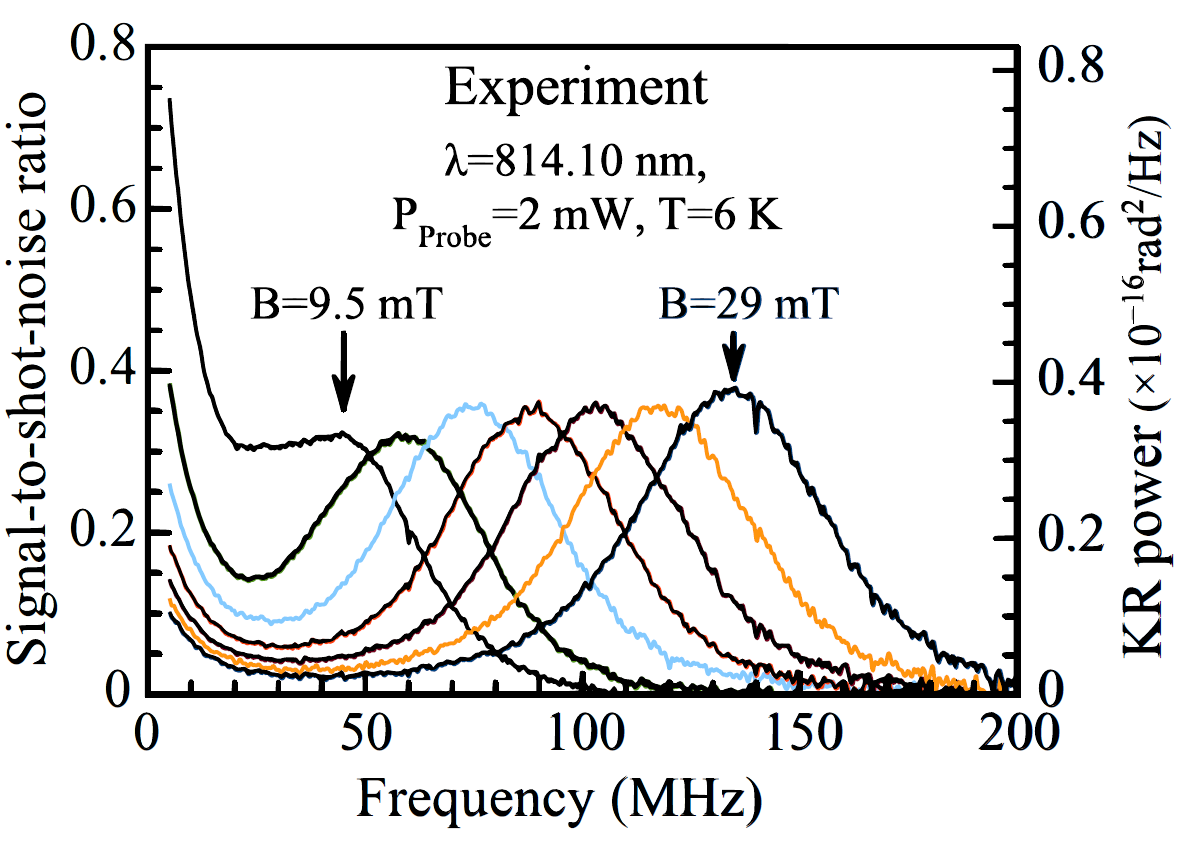}
	\caption{ Kerr rotation noise spectra in a single 20 nm wide GaAs/AlAs quantum well (placed in a microcavity) measured in magnetic fields ranging from 9.5 to 29 mT with equal increments. Experimental parameters are shown in the panel. Figure adapted from Ref. \cite{Poltavtsev}.}
	\label{Fig5} 
\end{figure}

Experiments on measuring stochastic fluctuations of electron spins were first performed by Aleksandrov and Zapasskii (1981), who used optical Faraday rotation to detect spin fluctuations in a gas of sodium atoms \cite{Aleksandrov}.
A similar optical method for measuring electron spin noise in condensed matter was demonstrated, in particular, in doped $n$-GaAs by M. Oestreich and his colleagues \cite{Oestreich2005}, see also \cite{Crooker}. Fig.~\ref{Fig5} shows the spin noise spectrum  measured in a single 20 nm wide GaAs/AlAs quantum well placed in a microcavity \cite{Poltavtsev}. 

At equilibrium, by virtue of the Callen-Welton fluctuation-dissipation theorem \cite{Landau5}, the spectrum of fluctuations of the rotation angle $\theta$ is proportional to the imaginary part of the magnetic susceptibility $\chi''$ in Ref.~(\ref{chi}) \cite{GlazovIvch2012}. Thus, there is a direct connection between the EPR and the spin noise method. A physical connection of these two phenomena with the spin-flip Raman scattering and spin beats (\ref{St}) can also be established  \cite{Zapasskii}.

\section{  Electron $g$ factor in bulk semiconductors} \label{bulk}

Experimental values of the $g$ factor of the electron in the conduction band of direct-gap semiconductors A$_3$B$_5$ or A$_2$B$_6$ with a zinc blende structure vary within the range from $-51.3$ in InSb and $-14.8$ in InAs to $-0.44$ in GaAs and $1.9$ in ZnTe. This wide range is well described by the formula called the Laura Roth formula  \cite{Roth}
\begin{equation} \label{gRoth}
	g =  g_0 - \frac43 \frac{|p_{cv}|^2}{m_0} \frac{\Delta}{E_g (E_g + \Delta)}\:.
\end{equation} 
Here $g_0$  is the free electron $g$ factor (\ref{g0}), $E_g$ is the band gap energy, $\Delta$ is the spin-orbit energy splitting of the valence band, $m_0$  is the electron mass in vacuum, $p_{cv}$  is the interband matrix element of the momentum operator. The formula (\ref{gRoth}) is obtained using the ${\bm k}\cdot {\bm p}$ perturbation theory method taking into account ${\bm k}\cdot {\bm p}$ admixture of states from the $\Gamma_8$ and $\Gamma_7$ valence subbands to the states of the $\Gamma_6$ conduction band. 

The formula (\ref{gRoth}) is analyzed in detail in the papers \cite{Hermann-exp,HermannPRB} and Chapter 11 of the collective monograph \cite{Hermann1977}. The analysis shows that the typical value of the interband interaction energy $E_p = 2 |p_{cv}|^2/m_0$ \cite{Agrawal1972,lawaetz1971} (or $P^2$ in the notation of \cite{HermannPRB})
is very similar for most binary compounds A$_3$B$_5$ or A$_2$B$_6$, in contrast to the wide scatter of $E_g$ and $\Delta$ values. Therefore, for estimates, $E_p$ can be set as a constant with a value of 20 eV, and the values of $E_g$ and $\Delta$ can be taken from optical experiments. A more accurate calculation involves determining the energy $E_p$ from the ${\bm k}\cdot {\bm p}$ formula for the effective mass $m^*$ when the value of this mass is found, for example, from the cyclotron resonance experiment. Finally, for even better agreement between theory and experiment, a small constant contribution $g_{\rm rb}$ can be added to the right-hand side of the equation (\ref{gRoth}), taking into account the ${\bm k}\cdot {\bm p}$ contribution of remote bands to the electron $g$ factor. We emphasize that the second term in \eqref{gRoth}, due to the spin-orbit interaction in the valence band, plays an important role and makes the main contribution to $g$. This contribution is greater, the greater the spin-orbit splitting $\Delta$ and the smaller the band gap $E_g$ are, and, for many semiconductors, leads to a change in the sign of the electron $g$ factor. For example, $g({\rm CdTe}) = - 1.66$ at $\Delta =0.82$ eV and $E_g=1.6$ eV, $g({\rm GaAs}) = -0.44$ at $\Delta =0.34$ eV and $E_g= 1.52$ eV, $g({\rm CdSe}) = 0.68$ at $\Delta =0.42$ eV and $E_g=1.8$ eV, and $g({\rm InP}) = 1.2$ at $\Delta =0.108$ eV and $E_g=1.42$ eV (see references in \cite{Semina2021}).

Recently, the Roth formula has been used to calculate the $g$ factor of holes in the valence band of lead-halide perovskites APbX$_3$ (X $-$ Cl, Br, I; A $-$ Cs, methylammonium MA, formamidine FA) \cite{NatCom2022}. Compared to GaAs-type semiconductors, the bands in these cubic symmetry materials are inverted: the valence band $R^+_6$ (analog of the $\Gamma_6$ band) is simple, and the conduction band consists of two subbands $R^-_6$ and $R^-_8$ (analogs of the $\Gamma_7$ and $\Gamma_8$ subbands) with a negative spin-orbit splitting $\Delta$. The absolute value of the spin-orbit splitting in lead halide perovskites is of the order of 1.5 eV, and the band gap can vary from 1.0 eV to 3.5 eV depending on the composition. Accordingly, the hole g factor varies in the range from -3 to +1.5 \cite{NatCom2022}.  

\section{Electron $g$ factor in quantum wells and other two-dimensional systems} 

The electron $g$ factor in a quantum well structure was first calculated in \cite{IcvhKis1992}, see also \cite{IcvhKis1998,IcvhKis1998a,Kiselev1999}. The calculation was carried out within the Kane model, which accurately takes into account the ${\bm k} \cdot {\bm p}$ mixing of the $\Gamma_6$, $\Gamma_7$, and $\Gamma_8$ bands. The electron wave function is sought in the form
\begin{equation} \label{Psi}
	\Psi = u_{\frac12}(z) \uparrow S + u_{-\frac12}(z) \downarrow S +
	\sum\limits_{i=x,y,z} \left( v_{i; \frac12}(z) \uparrow R_i + v_{i; -\frac12}(z) \downarrow R_i \right)\:,
\end{equation}
where $S({\bm r})$ and $R_i ({\bm r}) = X,Y,Z$ are the orbital Bloch functions at the bottom of the conduction band and the top of the valence band, constructed  neglecting the spin-orbit interaction, $\uparrow$ and $\downarrow$  are the spin columns, $u_{\pm 1/2}$ and $v_{i;\pm 1/2}$  are smooth envelope functions. A three-layer heterostructure is considered, consisting of a layer of material A (well) placed between semi-infinite layers of the barrier B, with the $z \parallel [001]$ axis being perpendicular to the plane of the interfaces. The smooth envelopes, represented as two-component columns $u$ and $v_i$, satisfy the normalization condition
\begin{equation} \label{uvnorm}
	\int d {\bm r} (u^\dagger u + {\bm v}^\dagger \cdot {\bm v}) =1 
\end{equation}
and equations 
\begin{eqnarray} \label{uv}
	&&E u=-{\rm i} P\: \hat{\bm k} \cdot {\bm v}  \:,\\&&
	\left(E+E_g+\frac\Delta3\right) {\bm v}=\:\:{\rm i} P\: \hat{ {\bm
			k} }u+ {\rm i} \frac\Delta3 \: {\bm \sigma}\times {\bm v} \:. \nonumber
\end{eqnarray}
Here $P = {\rm i} \hbar p_{cv}/m_0 = \hbar^2 \langle S| \nabla_z | Z \rangle/m_0 $, $\hat{\bm k} = - {\rm i} {\bm \nabla}$. The parameters $E_g, \Delta$ and $P$ are different for materials A and B, the electron energy $E$ is measured from the bottom of the conduction band of the corresponding material. Unlike a bulk crystal, to find the electron states in a heterostructure, it is necessary to introduce boundary conditions at the interfaces. In Ref.~\cite{IcvhKis1992} the boundary conditions 
\begin{equation} \label{BC}
	\left( u_{\pm \frac12}\right)_{z = - 0} = \left(u_{\pm \frac12}\right)_{z = + 0}\:,\:\left(P v_{z;\pm \frac12}\right)_{z = - 0} = \left(P v_{z;\pm \frac12}\right)_{z = + 0} \:
\end{equation}
were used, where $z = \pm 0$ are the limiting right and left sides of the interface between two materials. When $P_A=P_B$, these conditions transform into the Suris boundary conditions \cite{Suris}. This approximation corresponds to a small spread of the Kane energy $E_p = 2m_0 P^2 /\hbar^2$. 

There are several equivalent procedures for calculating the transverse and longitudinal $g$ factors
$g_{\perp}$ and $g_{\parallel}$, here we follow Ref.~\cite{Kiselev1999}. The transverse $g$ factor was calculated using the formula
\begin{equation} \label{g-work}
	\frac12 \mu_B \sigma_{i, ss'}g_{ij} B_j = \frac12\: g_0\: \mu_B\: \eta_{i; ss'} B_j + V_{ss'} \:,
\end{equation}
where
\begin{eqnarray} \label{g-perp}
	&&V_{ss'} = - {\rm i} \:\frac{e}{c \hbar}\: \int P\: \left[\: \left( {\bm A}
	{\bm v}^{+}_{s} \right) u_{s'} - u^{+}_{s} \left( {\bm A} {\bm
		v}_{s'} \right) \:\right] d {\bm r}\:, \\ && \eta_{i; ss'} = \langle e1,s|
	\sigma_i|e1,s'\rangle \:, \nonumber
\end{eqnarray}
the states $|e1,s \rangle$ are the solution $u_s(z),{\bm v}_s(z)$ \addELI{of} Eqs.~(\ref{uv}) with a given spin direction: $u_s (z) \neq 0$, $u_{-s} (z) \equiv 0$.	

To find the component $g_{\perp}$, one can take the vector potential ${\bm A}$ in the gauge $(0,-B_x z, 0)$ or $(B_y z,0, 0)$. Equation~(\ref{g-work}) is also applicable for calculating electron $g$ factors in quantum wires and quantum dots. In Ref.~\cite{IcvhKis1998a}, to find the longitudinal $g$ factor in a quantum well of width $L$, the component $g_{zz}$ was calculated in a quantum wire with a rectangular cross section $2b \times L$ with edges perpendicular to the $z$ and $x$ axes and in the gauge ${\bm A} = (0,B_z x,0)$, and then the limiting value of $g_{zz}$ was found at $b \to \infty$.

\begin{figure}[h!]
	\centering
	\includegraphics[scale=0.75]{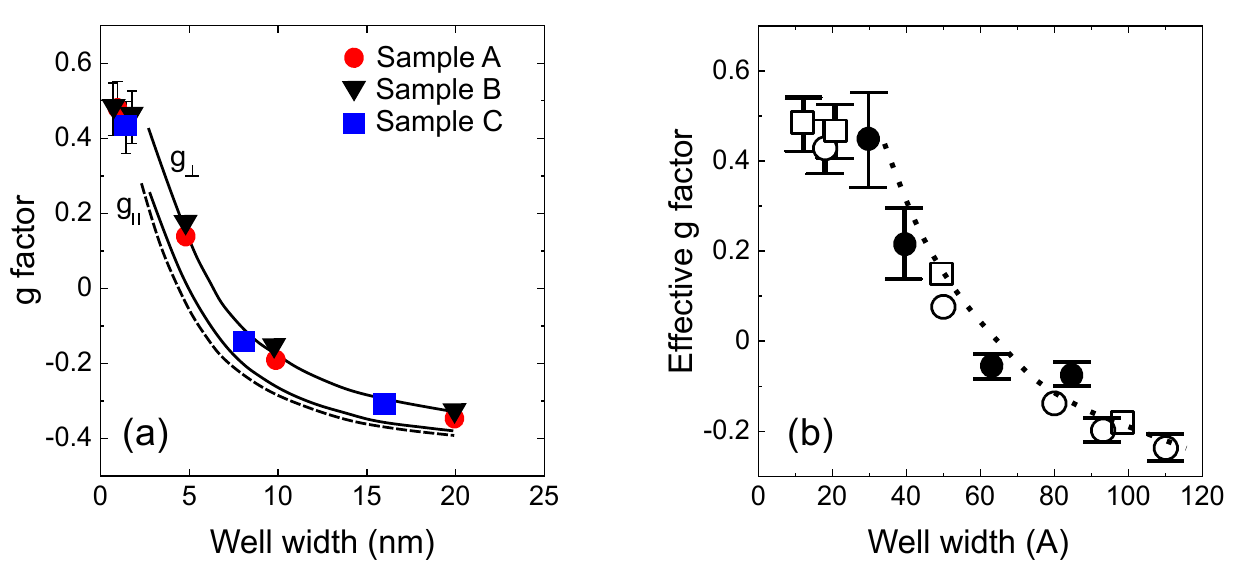}
	\caption{ Experimental results on the transverse electron $g$ factor measured  in a GaAs/Al$_x$Ga$_{1-x}$As quantum well structure as a function of the well width. (a) Data for three samples A ($x=0.35$), B ($x=0.3$) and C ($x=0.27$) \cite{Oestreich1995}. Dashed and solid lines show the results of theoretical calculations in the single-band approximation and in the Kane model from \cite{IcvhKis1992}. (b) Data for $x=0.3$ given in the paper \cite{Potemski1996} from the Refs.~\cite{Oestreich1995} (open squares), \cite{Potemski1996} (black circles) and \cite{Snelling1992} (open circles), the dotted line is the results of the theoretical calculation from Ref.~\cite{IcvhKis1992}.   }
	\label{Fig6} 
\end{figure}

Figures \ref{Fig6}(a) and \ref{Fig6}(b) present experimental data from Refs.~\cite{Oestreich1995,Potemski1996,Snelling1992} on the 
measurements of the electron transverse $g$ factor in a GaAs/Al$_x$Ga$_{1-x}$As quantum well structure as a function of the well width. In addition to the experimental data, the authors also presented theoretical curves calculated in Ref.~\cite{IcvhKis1992}. It is evident that good agreement was achieved between theory and experiment.

In Ref.~\cite{IcvhKis1992}, a noticeable anisotropy of the electron $g$ factors in a quantum well was predicted, the curves  $g_{\perp}$ and $g_{\parallel}$ in Fig. \ref{Fig6}(a) \cite{Oestreich1995}. This anisotropy was experimentally observed in a number of studies, see Refs.~\cite{Kiselev2007,Anisotropy1992,Kowalski1994}. In Ref.~\cite{IcvhKis1997}, it was theoretically shown that the longitudinal component of the electron $g$ factor is more sensitive to the application of an external electric field than the transverse one.

The Zeeman effect manifests itself in a peculiar way in monolayers of transition metal dichalcogenides $MX_2$ \cite{TMDC,DurnGlaz,Anvar}. In these two-dimensional materials of extremely thin thickness, the bottom of the conduction band and the top of the valence band are located at the vertices of the ${\bm K}_{\pm}$ hexagonal Brillouin zone. Moreover, the electron and hole states at these points are spin-split already in zero magnetic field. By $g$ factors we mean the parameters $g_{c,v}^{\downarrow {\bm K}_-} = g_{c,v}^{\uparrow {\bm K}_+}$, describing shifts of the charge carrier energy at the lower spin sublevel in the ${\bm K}_+$ and ${\bm K}_-$ valleys, which are linear in the external magnetic field. It was shown that the main contribution to the exciton $g$ factor $g_{X^0} = g_e - g_v$ comes from the remote bands: the upper conduction band $c+2$ and the lower valence band $v-3$ (in notations of Refs.~\cite{TMDC,DurnGlaz}, see Figs.~3, 4 and Table~1 in \cite{DurnGlaz}). Good agreement with experimental data for the WSe$_2$ monolayer was obtained in calculations by the density functional method (DFT) \cite{Anvar} for the $g$ factor of bright ($g_{\rm DFT}= -4.0$, $g_{\rm exp}= -4.1$) and dark ($g_{\rm DFT}=10.1$, $g_{\rm exp}=9.4$) excitons.

\section{Electron $g$ factor in quantum dots and nanocrystals} 

The size dependence of the $g$ factor of an electron localized in quantum dots and nanocrystals (NCs) was studied in Refs. \cite{IcvhKis1998,IcvhKis1998a,Gupta2002,Rodina2003,Prado2004,Delerue2017,Semina2020,
	Semina2021,Polish2020,Zhukov2025}. In Refs. \cite{IcvhKis1998,IcvhKis1998a}, Eq.~(\ref{g-work}) was also applied to calculate the $g$ factors in quantum wires and quantum dots. In spherical quantum dots, the conduction band component of the electron envelope  wave function in the ground state can be sought in the form
$u_s({\bm r}) = Y_{00} f(r) c_s$, where $c_s$ ($s=\pm 1/2$) describes the spin states, $Y_{00} =1/\sqrt{4\pi}$ is the spherical harmonic of the angular momentum $l=0$, and $f(r)$ is the real radial wave function \cite{IcvhKis1998a,Rodina2003}. As a result, the $g$ factor of an electron in a quantum dot with the spherical boundary between two materials $A$ and $B$ at $r=R$ is described  by
\begin{eqnarray} \label{gAB}
	g- g_0 = &&\left[g_A(E) -g_0 \right] w_A+ \left[g_B(E) -g_0 \right] w_B + \left[g_B(E) -g_A(E) \right] w_{\rm surf} \, , \\
	&& \, \, w_{A,B} = \int_{A,B} f^2(r) r^2 dr \, , \qquad w_{\rm surf}=  f^2(R) \frac{4\pi R^3}{3} \, . \nonumber
\end{eqnarray}
The first two terms on the right-hand side of Eq.~(\ref{gAB}) describe the volume contributions of regions A and B, taking into account the energy-dependent $g$ factors $g_{A,B}(E)$ according to
\begin{equation} \label{gE}
	g(E) =  g_0 - \frac{2E_p}{3} \frac{\Delta}{(E_g+E) (E_g + E+ \Delta)}\:.
\end{equation} 
Note that  $w_A+w_B <1$  due to the normalization condition (\ref{uvnorm}). The third term on the right-hand side of  Eq.~(\ref{gAB}) describes the interface contribution.  Equation (\ref{gAB}) can be adapted to calculate the electron $g$ factor in cylindrical quantum wires as well as the transverse $g$ factor in quantum wells.

\begin{figure}[h!]
	\centering
	\includegraphics[scale=0.5]{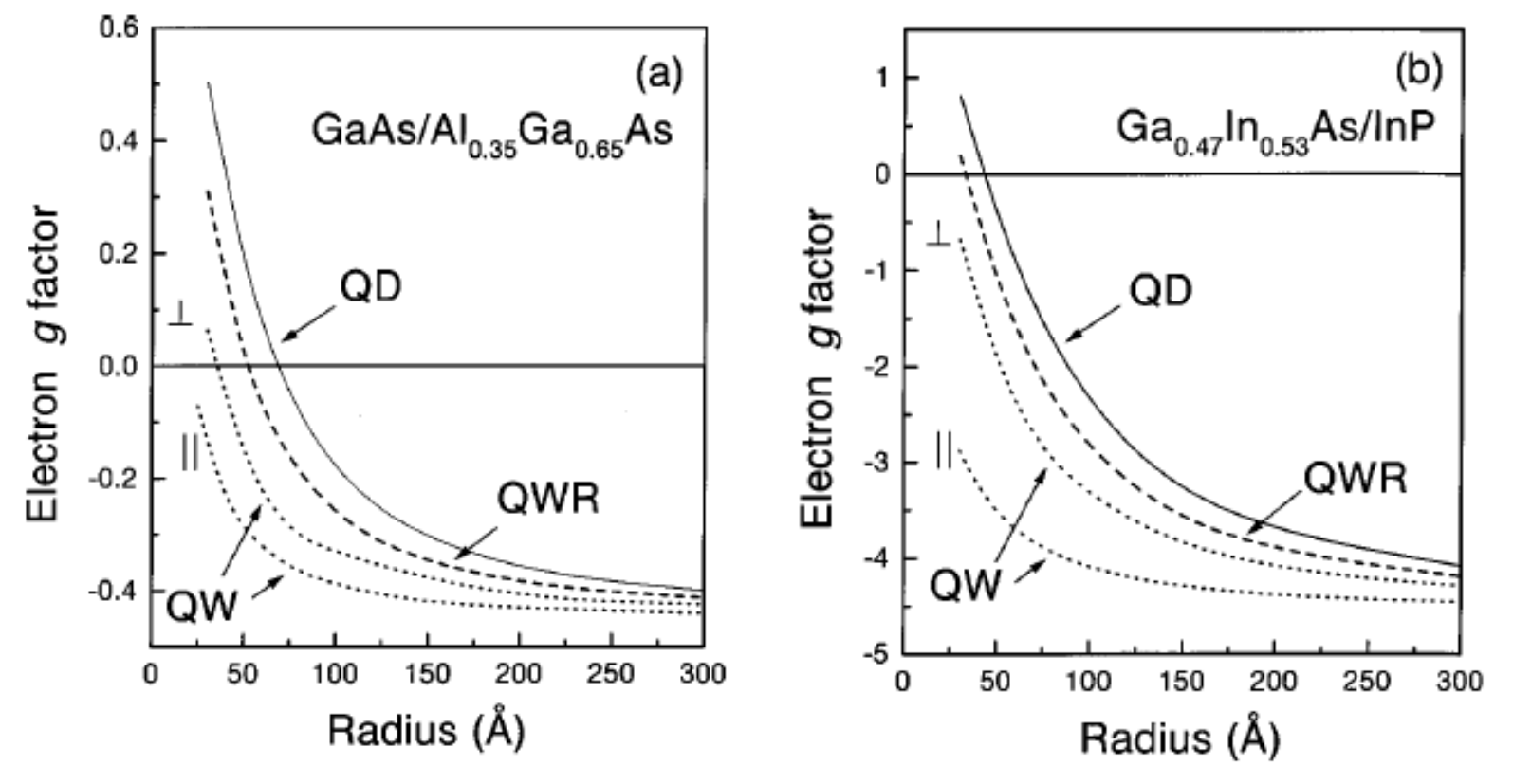}
	\caption{Dependence of the electron $g$ factor on the radius $R$ in spherical quantum dots [solid line, Eq.~(\ref{gAB})], cylindrical quantum wires (dashed line), and quantum wells
(dotted line, $R=L_z/2$) for heterostructures (a) GaAs/Al$_{0.35}$Ga$_{0.65}$As  and (b) Ga$_{0.47}$In$_{0.53}$As/InP. Figure from Ref. \cite{IcvhKis1998a}. }
	\label{Fig7} 
\end{figure}

Figure \ref{Fig7} shows the size dependencies of the electron $g$ factor in  spherical quantum dots, cylindrical quantum wires, and quantum wells calculated in Ref.~\cite{IcvhKis1998a}. In the limit of large radii $R$, the $g$ factor values tend to the bulk value $g_A$. For better agreement, the value $g_{\rm rb}$, corresponding to the contribution of remote bands in the bulk semiconductor A, was added to $g$.

In spherical semiconductor NCs dispersed in a glass matrix or synthesized in a colloidal solution, the potential barriers at the surface are often high enough to be impermeable for the electron density, which corresponds to the zero probability density flux, $J_\tau(R)=0$, across the spherical boundary. Equation (\ref{gAB}) can be applied to calculate the size dependence of the electron $g$ factor in NCs made of material $A$ by setting $g_B(E) \equiv g_0$.

When the boundary condition $f(R)=0$ is applied, the interface (or surface) contribution to the electron $g$ factor vanishes. However, the condition $f(R)=0$ is not the only possible one that guarantees $J_\tau(R)=0$ for an impenetrable barrier. The point is that the valence band component ${\bm v}(R) \propto df/dr \ne 0$. In the article \cite{Rodina2003}, general boundary conditions were proposed that guarantee $J_\tau(R)=0$ and are characterized by the surface parameter $A_e$. The parameter $A_e$ has the dimension of length and characterizes the surface layer. In this case, the surface contribution $g_{\rm surf} \ne 0$ and its value as well as the size dependence of the electron $g$ factor as a whole, depend on the value of $A_e$. Modeling of the dependence of the electron $g$ factor at the lowest quantum confinement level on the radius of colloidal CdSe NCs, which was measured in Ref.~\cite{Gupta2002} using the Faraday rotation signal in the pump-probe method, allowed the authors to determine the optimal value of the surface parameter $A_e = -0.06$~nm \cite{Gupta2002,Rodina2003}. The presence of a surface contribution to the electron magnetic moment is due to additional spin-orbit interaction in the conduction band induced by the admixture of valence band states and the presence of a spherical boundary.	

\begin{figure}[h!]
	\centering
	\includegraphics[scale=0.2]{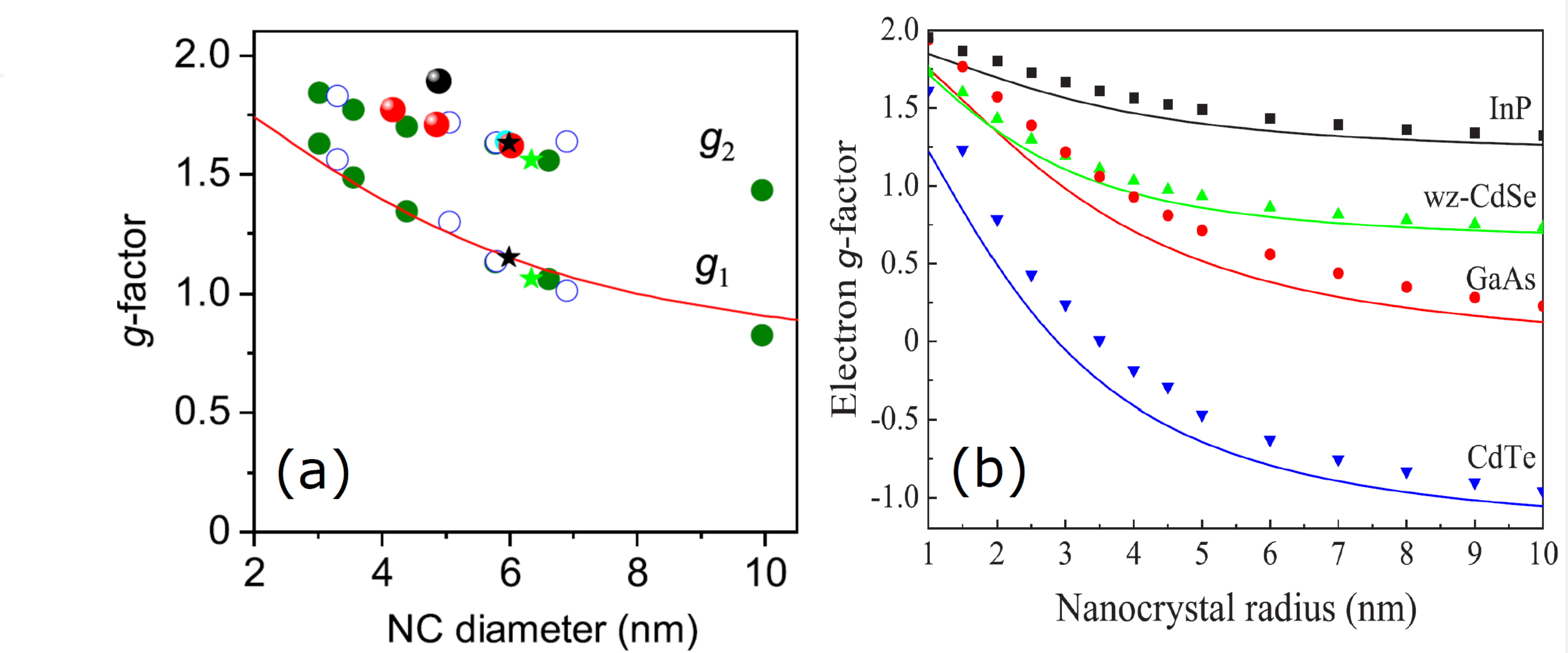}
	\caption{Dependence of the electron $g$ factor on the diameter (a) or radius (b) of a spherical nanocrystal. (a) Symbols show experimentally measured values in CdSe NCs collected in Ref. \cite{Zhukov2022} from various studies, including \cite{Gupta2002} (open circles) and \cite{Hu2019} (green circles) for colloidal CdSe and \cite{Zhukov2022} (red and black circles) for CdSe in glass. The solid line shows the result of a theoretical calculation using Eq.~(\ref{gAB}) for the CdSe parameters from \cite{Semina2021}. (b) Comparison of the results of calculation using Eq.~(\ref{gAB}) with the results of calculation in the tight-binding model \cite{Delerue2017} for different materials with the same band parameters \cite{Semina2021}. }
	\label{Fig8} 
\end{figure}

The value of $A_e$ determined in Refs. \cite{Gupta2002,Rodina2003} is additional to the set of CdSe bulk parameters used in calculations: $E_p, E_g, \Delta, g_{\rm}$. Later, additional studies of the size dependence of the electron $g$ factor in CdSe NCs were carried out. The measurement results from various studies are collected in \cite{Zhukov2022} and shown by symbols in Fig.~\ref{Fig8}(a). It is seen that the $g$ factor $g_1$ of an electron at the lowest quantum confinement energy level (ground state) is well described by Eq.~(\ref{gAB}) with $g_{\rm surf}=0$ for the CdSe parameters used in Ref. \cite{Semina2021}. The nature of the second $g$ factor $g_2$ with a larger value experimentally observed in the pump-probe method is not precisely known. Since the first observation in Ref.~\cite{Gupta2002}, various hypotheses have been put forward and tested. In particular, it was shown experimentally \cite{Hu2019} and theoretically \cite{Zhukov2022} that $g_2$ cannot be attributed to the  exciton $g$ factor. Moreover, it was shown theoretically \cite{Golovatenko2025} that even in the presence of anisotropy, $g_\parallel \neq g_\perp$, in wurtzite or spheroidal NCs, with their arbitrary orientation in the ensemble, the observed Larmor precession frequency is always determined by the transverse $g$ factor $g_\perp$. Currently, it is assumed that the value of $g_2$ corresponds to an electron additionally localized near the surface \cite{Hu2019,Zhukov2022}. The nature of such localization, however, remains unclear. It is interesting to note that while both electron $g$ factors, $g_1$ and $g_2$, were observed in colloidal CdSe NCs, only the value of $g_2$ was observed in CdSe NCs in a glass matrix, both in the pump-probe method in Ref.~\cite{Zhukov2022} and the spin-flip Raman scattering experiment in a recent work \cite{Ina2025}.

Figure \ref{Fig8}(b) shows a comparison of the dependencies of the electron $g$ factor at the lowest quantum confinement level on the NC  radius calculated in the tight-binding method in Ref.~\cite{Delerue2017} and according to Eq.~(\ref{gAB}), for different materials with the same band parameters in both approaches and for $g_{\rm surf}=0$ \cite{Semina2021}. Evidently, the calculation using the eight-band ${\bm k} \cdot {\bm p}$ model is in good agreement with the results of atomistic calculations.

 In the recent work \cite{Zhukov2025}, an extension of the Roth formula was proposed for calculating the dependence of the electron $g$ factor on the NC radius within the second-order perturbation theory, taking into account the interaction of the electron in the ground state with the quantum-confined states of holes in the valence subbands $\Gamma_8$ and $\Gamma_7$. The interaction not only with the ground but also with the excited states of holes becomes possible due to the allowance for the complex structure of the valence band within the six-band Luttinger Hamiltonian  \cite{Luttinger1956}. The calculations \cite{Zhukov2025} were carried out for CuCl NCs in which $\Delta < 0$. Nevertheless, the developed generalization is also applicable for semiconductors with $\Delta>0$.

Interestingly, that despite more than forty years of research of CuCl NCs in a glass matrix, which gave a start to the history of quantum dots, the size dependence of the electron $g$ factor has remained largely unexplored until recently.
In recent magneto-optical studies of CuCl NCs in a glass matrix, the Larmor frequency (obtained from the ellipticity signal in the pump-probe method) and the Stokes shift energy in spin-flip Raman scattering were measured as functions of the excitation energy and, thus, of the NC radius. It is unknown, however, to which resident charge carrier (electron or hole) or exciton the observed dependence corresponds. The CuCl nanocrystals were not intentionally doped, and resident carriers of both types could be created during photocharging. The measured $g$ factor is close in order of magnitude to the electron $g$ factor in bulk CuCl, the value of which $2.03 > 2$ is due to the negative value of the spin-orbit splitting of the valence band, $\Delta < 0$. The contribution of the valence band to the electron $g$ factor is parametrically small because of the large value of $E_g$ and decreases with increasing electron quantum confinement energy. Thus, both the Roth formula and the refined theory predict a decrease in the $g$ factor of the electron with decreasing NC radius, whereas in the experiment an increase of the $g$ factor with increasing energy is observed \cite{Zhukov2025}.

\section{Hole $g$ factor}

The valence band of elementary semiconductors Si, Ge (diamond crystal structure, O$_h$ point group symmetry) and A$_3$B$_5$ or A$_2$B$_6$ binary semiconductors (zinc-blende structure, cubic symmetry point group T$_d$) is formed from atomic $p$-orbitals and it is characterized by the orbital angular momentum $I=1$. Spin-orbit interaction leads to splitting of the complex valence band into the upper $\Gamma_8^+$ subband (or $\Gamma_8$ in the case of T$_d$ symmetry) and the spin-orbit split-off $\Gamma_7^+$ subband (or $\Gamma_7$).

The hole state  at the top of the valence subband $\Gamma_8^+$ or $\Gamma_8$ is fourfold degenerate, and the effective hole Hamiltonian has the form of a matrix of dimension 4$\times$4. The Zeeman Hamiltonian for holes in such a band, first proposed by J.M. Luttinger in 1956 \cite{Luttinger1956}, is characterized by two constants and has the form \cite{Bir_rus_book,IvchenkoPikus}
\begin{equation}\label{Hzh1}
	\widehat{H}_{Z}^{(\Gamma_8)}= 2 \mu_B  \left[ \mathcal{k}  {\bm J}{\bm B} + \mathcal{q} \left( J_x^3 B_x + J_y^3 B_y + J_z^3 B_z \right) \right]\:,
\end{equation}
where $J_i~(i=x,y,z)$ are matrices of angular momentum $J=3/2$. Constants with opposite signs $\varkappa = -\mathcal{k}$ and $q = - \mathcal{q}$ are often used in the literature following Ref.~\cite{Luttinger1956}.

The dimensionless parameter $\varkappa$ and the Luttinger parameters $\gamma_1, \gamma_2, \gamma_3$ were calculated within the framework of ${\bm k}$$\cdot$${\bm p}$ perturbation theory in Refs.~\cite{Dresselhaus1955,Roth}. Each of them was expressed as a linear combination of four parameters $F,G,H_1,H_2$, contributions to which are made, respectively, by the bands $\Gamma_2^-, \Gamma_{12}^-, \Gamma_{15}^-, \Gamma_{25}^-$ in the O$_h$ point group or $\Gamma_1, \Gamma_3, \Gamma_5, \Gamma_4$ in the T$_d$ point group. Since even in the thirty-zone ${\bm k}$$\cdot$${\bm p}$ model the $\Gamma_{25}^-$ or $\Gamma_4$ zone is absent \cite{Dresselhaus1955,Richard2004}, the parameter $H_2$ can be safely set equal to zero.
By inverting the relation between the parameters $\gamma_1, \gamma_2, \gamma_3$ and $F,G,H_1$, we can express the coefficient $\varkappa$ in terms of the Luttinger parameters \cite{Semina2023}:
\begin{equation}\label{kappa}
	\varkappa = -{\cal k} =  -\frac13\left(2 + \gamma_1 - 2\gamma_2 - 3\gamma_3 \right) \:.
\end{equation} 
Note that taking into account only bands with the $\Gamma_1$ and $\Gamma_5$ symmetry  in ${\bm k}$$\cdot$${\bm p}$ method results in the relation $\gamma_1 = 3 \gamma_3 - \gamma_2 -1$  and simplification of the expression (\ref{kappa}): $\varkappa = \gamma_2 - \frac13$.

The second magnetic Luttinger parameter $q=-{\cal q}$ describes the anisotropy of the Zeeman effect allowed by the cubic symmetry of the $O_h$ or $T_d$ point groups. In the ${\bm k}$$\cdot$${\bm p}$-theory, it arises only due to the spin-orbit admixture of states from remote bands to the states of $\Gamma_8$ valence subband, and can usually be neglected \cite{Marie1999, Winkler}. In this case, the Zeeman splitting of the hole does not depend on the direction of the magnetic field and can be described by an isotropic effective Hamiltonian
\begin{equation}\label{HzhG8}
	\widehat{H}_{Z}^{(\Gamma_8)}= - 2 \varkappa \mu_B\ {\bm J} {\bm B}.
\end{equation}

The quantum-confined states of a hole  in a nanostructure differ significantly from those in the bulk crystal. For spherically symmetric structures and in the  isotropic approximation for the Luttinger Hamiltonian ($\gamma_2 = \gamma_3 \equiv \gamma$), a good quantum number is the hole total momentum $ \bm{\mathcal{J}} = {\bm J}+ {\bm l}$, where ${\bm l} = {\bm r} \times {\bm k}$ is the hole dimensionless orbital momentum; $\bm r$ and $\bm k$ are the radius vector and wave vector of the hole, respectively. In this case, the Zeeman splitting of the hole is also independent of the external magnetic field  direction,  and it can be described by the isotropic effective Hamiltonian \cite{Efros1996,EfrosCh3}
\begin{equation}\label{Hzhsph}
	\widehat{H}_{Z}^{\rm sph}= - g_h^{\rm sph} \mu_B\ { \bm{\mathcal{J}}} {\bm B}.
\end{equation}

For a hole localized in a structure of uniaxial symmetry, the only conserved  quantum number  is the projection  of total momentum $\bm{\mathcal{J}}$ onto the structure $z$ axis. In this case, for the direction of the magnetic field along or perpendicular to the $z$ axis, the Zeeman splitting $\Delta E_M = E_M- E_{-M}$ of heavy ($M \equiv \mathcal{J}_z =\pm 3/2$) and light ($M =\pm 1/2$) holes is described by effective $g$ factors depending on $|M|$. In particular, the longitudinal $g$ factor (for a magnetic field along the $z$ axis) is defined as
\begin{equation}\label{g_hM}
	g_{|M|}^\parallel = - \frac{\Delta E_M}{2M\mu_B B}.
\end{equation}
The definitions of the hole $g$ factor according to Eqs.~\eqref{Hzhsph} and \eqref{g_hM} are often used in the physics of colloidal nanocrystals \cite{Efros1996,EfrosCh3,Semina2021,Semina2023}. Note,
however, that the definition of the hole $g$ factor with the opposite sign is often used, see,~e.g.,~Refs.~\cite{Marie1999,holeKiselev,Rodina2001f,vanKesteren1990}. Moreover, for structures with a large energy splitting between heavy and light hole states, such as quantum wells, the Zeeman splitting of heavy holes is often defined as $\Delta E_{3/2} = E_{+3/2}-E_{-3/2}= g_{hh} \mu_B B_z$, where $g_{hh}=-3g_{3/2}^\parallel$ \cite{holeKiselev}. All these definitions of the hole $g$ factor are used to describe the same dependence of the hole Zeeman splitting in the magnetic field, but one should be careful when comparing the values of the hole $g$ factors given in different literature sources.

For  a hole confined in a nanostructure, states with different angular momentum projections onto the magnetic field direction are mixed, resulting in a significant renormalization of the Zeeman contribution \eqref{HzhG8} to the effective Hamiltonian of the localized hole. In addition, for localized holes, another, orbital, contribution  to the Hamiltonian in an external magnetic field appears. The perturbation describing this contribution can be obtained by replacing the hole wave vector ${\bm k}$ in the Luttinger Hamiltonian for the hole kinetic energy by ${\bm k} -\frac{e}{c}\bm A$, where $\bm A$ is the vector potential of the external magnetic field, and by keeping the terms linear in the magnetic field. This perturbation is given explicitly in Refs.~\cite{Semina2021,Semina2023}. The resulting Zeeman splitting of a localized hole depends significantly on the shape of the nanostructure.

In a spherically or cubically symmetric system, at zero magnetic field, the ground state of a hole from the $\Gamma_8$ valence subband is fourfold degenerate. Within the isotropic approximation, the wave function of the lowest-energy even state of a hole with total angular momentum $\mathcal{J}=3/2$, which  is in most cases the ground state, has the form \cite{Gelmont1971,JOSA1993}
\begin{equation}
	\Psi_{M} = 2\sum_{l=0,2} (-1)^{l-3/2+M} (i)^l R_{l}(r) \times\sum_{m+\mu = M}
	\left(
	\begin{array}{ccc}
		l & 3/2&\mathcal{J} \\ m&\mu&-M
	\end{array}
	\right)
	Y_{l,m} u_{\mu} \, .
	\label{Gelmont_functions}
\end{equation}
Here $Y_{lm}$ are spherical harmonics \cite{Edmonds, Varshalovich}, $\left(_{m~n~p}^{i~~k~~l}\right)$ are  Wigner's $3j$ symbols, $u_\mu$ are Bloch functions in $\Gamma_8$ valence subband with the spin projection $\mu$ on the $z$ axis \cite{Bir_rus_book}, $R_{l}(r)$ are the hole radial wave functions, their specific form depends on the type of localizing or quantum confining potential and band structure parameters. In Ref.~\cite{Gelmont1973}, an expression for the hole ground state $g$ factor in the case of a hole localized on a shallow acceptor was obtained. We present this expression in the form valid for a hole localized in any spherically symmetric potential, including spherical core/shell heterostructures \cite{Semina2021}:
\begin{equation}\label{Gelmontgen}
	g_{\rm h}^{\rm{sph}} =2\varkappa+\frac{4}{5}I_1^g+\frac{4}{5}I_2^g(\gamma_1-2\gamma-2\varkappa),  \end{equation}
\begin{eqnarray}I_1^g=\frac{1}{2}\int r^3dr\left(R_2\frac{dR_0}{dr}-R_0\frac{dR_2}{dr}-\frac{3}{r}R_0R_2 \right)  \:,\:
	I_2^g=\int r^2drR_2^2\:. \nonumber \end{eqnarray}
Here $\gamma=(2\gamma_2+3\gamma_3)/5$ corresponds to the isotropic approximation for the Luttinger Hamiltonian. The expression \eqref{Gelmontgen} allows one to calculate the isotropic $g$ factor of a hole after finding the radial wave functions in a given spherically symmetric potential. In this case, the $g$ factor $g_{\rm h}^{\rm{sph}}$ calculated by Eq.~\eqref{Gelmontgen} is independent of the localization region size. This is a general property of the four-band Luttinger model: the hole $g$ factor calculated within this model  depends only on the symmetry and type of the localizing potential, but not on the linear dimensions of the localization region \cite{Semina2015}. 

The minimal Hamiltonian that allows one to describe the size dependence of the hole $g$ factor  is a six-band Hamiltonian that simultaneously takes into account the states in the valence subbands $\Gamma_8^+$ and $\Gamma_7^+$ (or $\Gamma_8$ and $\Gamma_7$) \cite{Bir_rus_book,Rodina2001f,Semina2021}. 
As a matter of fact, the four-band model describes well only states with small wave vectors ${\bm k}$ near the top of the valence subband $\Gamma_8$. However, as  the size of  the localization region  decreases, e.g., the size of a nanocrystal, the hole size-quantization energy $E_h$ and values of the wave vectors $k_h$ in the hole wave function increase. At energies $E_h$ comparable to the spin-orbit splitting of the valence band $\Delta$, the admixture of states from the spin-split subband $\Gamma_7$ to states from the upper subband $\Gamma_8$ becomes significant and results in a size dependence of the  hole $g$ factor. In this case, a sum of the hole spin-orbit and Zeeman Hamiltonians, involving contributions of both valence subbands, has the form \cite{Luttinger1956}:
\begin{equation}\label{Hzh15}
	\widehat{H}_{\rm so}+	\widehat{H}_{Z}^{(\Gamma_8 + \Gamma_7)}= - \frac23 \Delta ({\bm s}{\bm I}) + g_0 \mu_B({\bm s} \bm B)-(1+3\varkappa)\mu_B({\bm {I}} \bm B).
\end{equation}
Here ${\bm s} = {\bm \sigma}/2$, ${\bm I} = (I_x,I_y,I_z)$, $I_{\alpha}$ are the angular momentum  $I=1$ matrices of dimension 3$\times$3, the first term describes the spin-orbit interaction, the second term is the spin contribution to the Zeeman splitting, similar to the contribution for electrons (\ref{HZ}), and the third term is the contribution of the internal orbital momentum.
In Ref.~\cite{Semina2021}, the expression \eqref{Gelmontgen} for the $g$ factor describing the splitting of a hole with a total angular momentum $\mathcal{J}=3/2$, where $\bm{\mathcal{J}}= {\bm s} + {\bm I}+{\bm l}$, is generalized for a six-band model with the positive spin-orbit splitting $\Delta >0$. A similar expression for the hole $g$ factor describing the splitting of a doubly degenerate hole state with $\mathcal{J}=1/2$ (for negative $\Delta$, when the $\Gamma_7$ subband is at the top of the valence band) is given in Ref.~\cite{Zhukov2025}.

\begin{figure}[h!]
	\centering
	\includegraphics[scale=0.5]{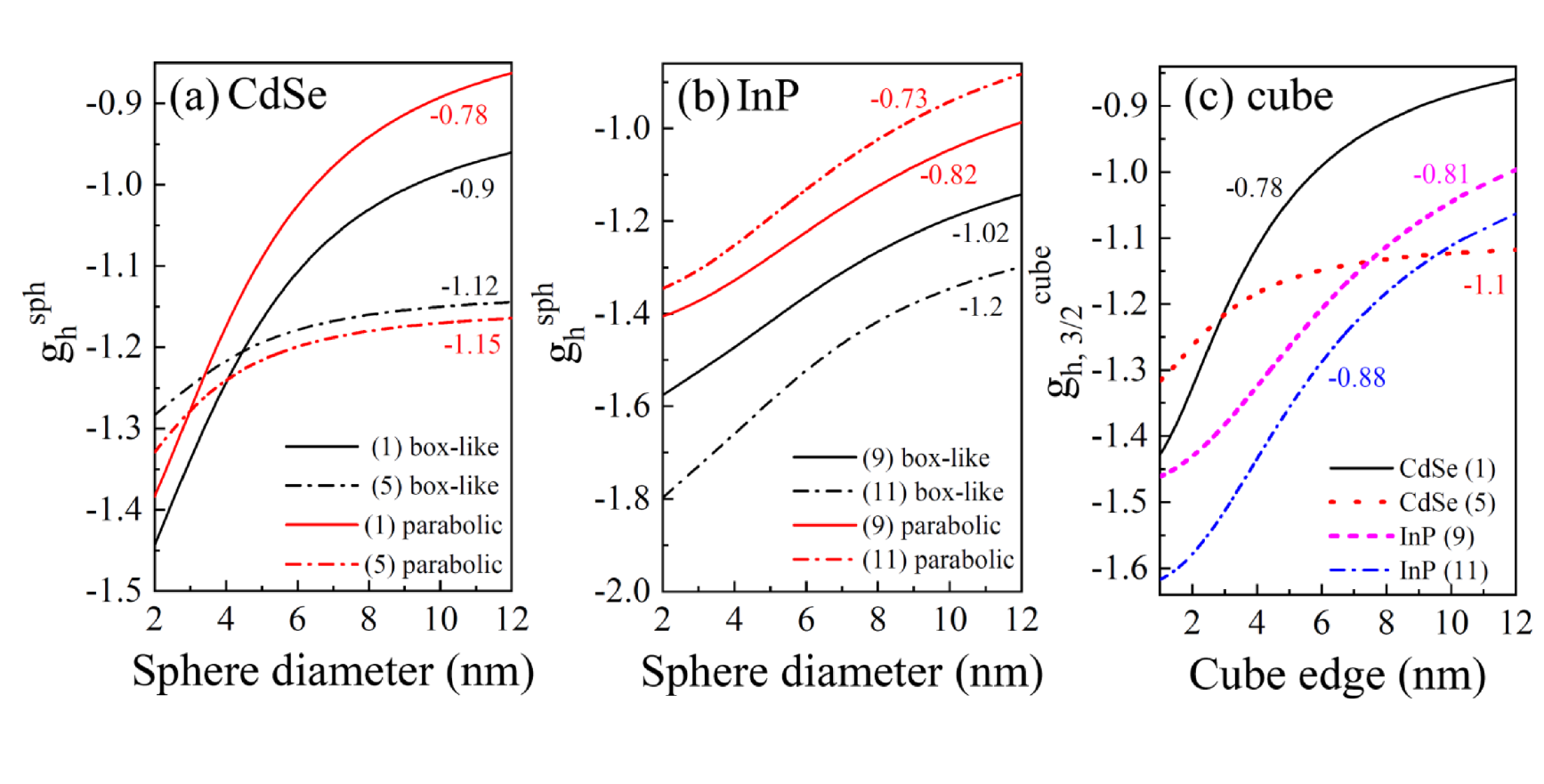}
	\caption{Dependence of the hole $g$ factor on the linear size of a spherical NC (a,b) with a box-like infinite potential (solid curves) and a smooth parabolic potential (dash-and-dot curves), as well as a cube NC (c) with a box-like potential. The numbers in brackets indicate rows in Table I of Ref. \cite{Semina2021}, where the parameters of the six-band Hamiltonian for CdSe and InP are listed. The numbers next to the curves indicate the asymptotic values of the hole $g$ factor in large-size NCs which can be calculated within the framework of the four-band Luttinger Hamiltonian. Figure is taken from Ref.~\cite{Semina2021}.}
	\label{Fig9} 
\end{figure}	

Figure~\ref{Fig9} shows an example of calculation of the dependence of the hole $g$ factor on the linear size of a NC of spherical, panel (a,b), or cube, panel (c), shape \cite{Semina2021}. For spherical NCs, two types of localizing potentials are considered: a box-like potential with an infinitely high barrier and a smooth parabolic potential. All the panels  of Fig.~\ref{Fig9} show a significant size dependence of the hole  ground state $g$ factor, which appears from different kinds of admixture of states from the spin-orbit split-off valence subband. Moreover, the smaller the value of $\Delta$ (for example, in InP $\Delta \approx 0.1$ eV), the larger the nanocrystal sizes at which asymptotic values valid for hole states with quantum-size energies much smaller than $\Delta$ are achieved. Interestingly, the asymptotic values of $g$ factors of the holes localized in large spherical and cube NCS lie around a value of $-1$ \cite{Semina2020,Semina2021}. For other nanostructures, such as quantum wires, quantum wells, and planar NCs, the dependence of the $g$ factor on the band parameters and the type of the localizing potential is more significant, even up to a change of sign \cite{Semina2020}.

For a hole confined in a semiconductor nanostructure, in addition to the isotropic contribution to the $g$ factor, there is also an anisotropic contribution, the symmetry of which is determined by the symmetry of the system. In general, the Zeeman effect can be described by second- and fourth-rank tensors.
\begin{equation}\label{effect_Z1}
	\hat{H}_Z^{\text{cube}} = - g^h_{\alpha \beta }\mathcal{ J}_\alpha B_{\beta} -2\mu_B Q_{ \alpha \beta \gamma \delta }\mathcal{ J}_\alpha \mathcal{ J}_\beta \mathcal{ J}_\gamma B_{\delta}\:,
\end{equation}	
where the tensor  $Q_{\alpha \beta \gamma \delta}$ is symmetric with  respect to the permutations of the first three indices. 
In the maximally low-symmetry structure (point group C$_1$) it has 30 linearly independent components. Since the sum of the squares $\mathcal{ J}^2_x + \mathcal{ J}^2_y + \mathcal{ J}^2_z $ for states with a given $\mathcal{ J}$ is the identity matrix multiplied by the scalar $\mathcal{ J} \left(\mathcal{ J} + 1\right)$, 21 linearly independent components of the tensor ${\bm Q}$ remain with the tensor ${\bm g}^h$ being renormalized. In a structure with the symmetry point group $O_h$ or $T_d$, the first tensor is diagonal: $g^h_{\alpha \beta } = g^h_{zz} \delta_{\alpha \beta}$, and the contribution of the fourth-rank tensor is reduced to a linear combination of the matrices $\bm{\mathcal{J}} {\bm B}$ and $\sum_{\alpha}\mathcal{J}^3_{\alpha} B_{\alpha}$. Then the expression \eqref{effect_Z1} is reduced to
\cite{Bir_rus_book,Semina2023,Malyshev2000}
\begin{equation}\label{effect_Z}
	\hat{H}_Z^{\text{cube}}= - g_{\rm h}\mu_B \left(\bm{\mathcal{J}}\bm B \right)-2 Q^{\text{eff}}\mu_B \left(\mathcal{ J}_x^3B_x+\mathcal{ J}_y^3B_y+\mathcal{ J}_z^3B_z\right)\:.
\end{equation}
In addition to the small crystallographic contribution $\propto q$, the ``source'' of the cubic anisotropy can also be the cubic anisotropy of the hole dispersion, the so-called valence band warping characterized by the parameter $(\gamma_2-\gamma_3)/\gamma$, and the cubic shape of the nanostructure. In this case, even for a spherically symmetric NC, the parameter $g_h$, which describes the isotropic contribution, differs from the parameter $g_h^{\rm sph}$ of the isotropic model (\ref{Gelmontgen}), since it contains a correction due to the cubic perturbation $\propto (\gamma_2-\gamma_3)/\gamma$.

In Ref. \cite{Semina2023} it was shown that the cubically-symmetric contributions to the hole  $g$ factor (parameter $Q^{\text{eff}}$), associated with the valence band warping in spherical and cube NCs, as well as those associated with the shape in cube NCs grown along one of the crystallographic axes of the crystal, can be comparable with the isotropic part proportional to $g_{\text{h}}$. Previously, a similar and comparable in magnitude contribution to $Q^{\text{eff}}$ caused by the valence band warping was obtained in  Ref. \cite{Malyshev2000} for a hole bound on an acceptor in cubic semiconductors.
\begin{figure}[h!]
	\centering
	\includegraphics[scale=0.6]{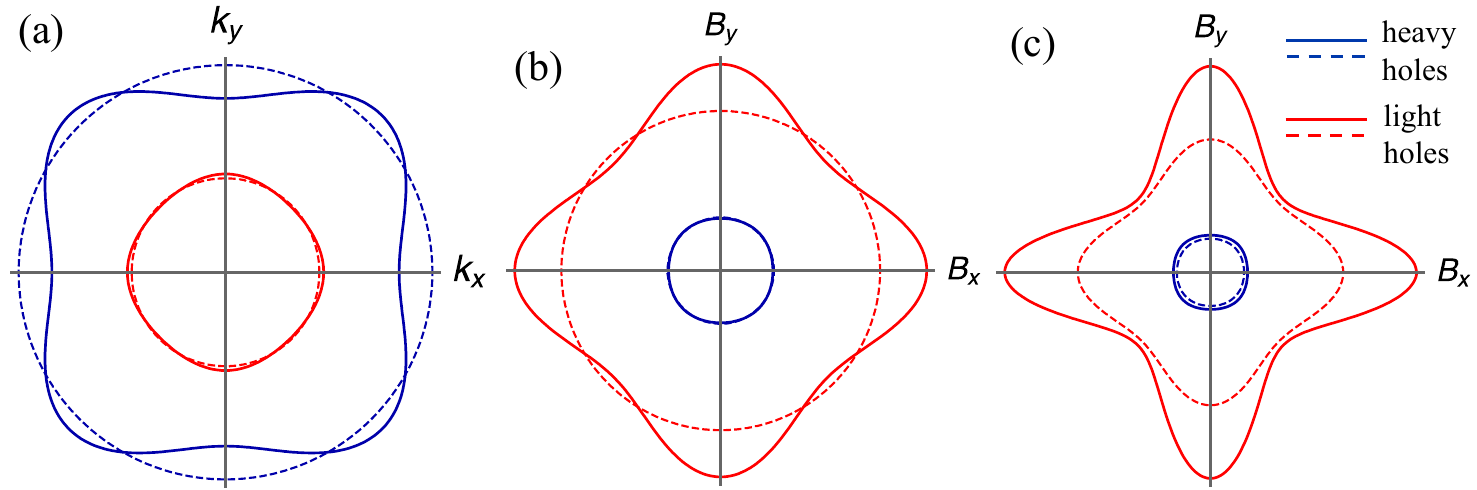}
	\caption{(a) Constant energy surfaces  of   free heavy holes (blue curves) and light holes (red curves) in the $(k_x, k_y)$ plane at $k_z = 0$ in the bulk crystal. (b) and (c) Surfaces of constant Zeeman splitting for the states of heavy (blue curves) and light (red curves) holes calculated for spherical and cube NCs with a box-like  potential. The calculation is performed for the CdSe parameters: $\gamma_1=2.52, \gamma_2=0.65, \gamma_3=0.95$, the dashed curves correspond to the isotropic approximation for the Luttinger Hamiltonian with $\gamma_2=\gamma_3=0.83$. The figure is adapted from Ref. \cite{Semina2023}.}
	\label{Fig10} 
\end{figure}

Figure~\ref{Fig10} illustrates the effect of the cubic symmetry on the hole $g$ factor in spherical and cube CdSe NCs \cite{Semina2023}. Panel (a) shows the constant-energy surfaces of bulk heavy and light holes in the $(k_x, k_y)$ plane at $k_z = 0$, where the $z$ axis is directed along one of the crystallographic axes $\langle 100 \rangle$; see, for example, Eq. (24.13a) in the book~\cite{Bir_rus_book}. Panels (b) and (c) show the surfaces of constant Zeeman splitting,
see, for example, Eqs. (39.21a) and (39.21b) in \cite{Bir_rus_book}, for hole states in spherical and cube NCs with a box-like potential. The solid curves are calculated taking into account the valence band warping, while the dashed curves correspond to the spherical approximation for the Luttinger Hamiltonian. It is evident that in spherical NCs (panel (b)), the curves calculated in the spherical approximation for the Luttinger Hamiltonian are circular, while the cubic anisotropy appears when the valence band warping is taken into account. For cube NCs, panel (c), anisotropy already exists even when using the isotropic approximation for the Luttinger Hamiltonian. As shown in Ref.~\cite{Semina2023} the contributions to $Q^{\text{eff}}$ determined by the NC shape and the valence band warping can be comparable and have the same or opposite signs, depending on the material band parameters.   

In the presence of additional uniaxial anisotropy, as for example along the $z$ axis, which coincides with the $[001]$ crystallographic direction (point group $D_{4h}$ or $D_{2d}$), the $z$ and $x,y$ axes become nonequivalent. In this case, the states of heavy and light holes are already split in zero magnetic field. The effective Hamiltonian \eqref{effect_Z} takes the form
\begin{eqnarray} \label{effect_Z_D4h}
	&&	\hspace{2.5 cm} \hat{H}_Z^{\text{eff}}=  -\mu_B \left[g_{\perp}(\mathcal{J}_xB_x+\mathcal{J}_yB_y)+g_{\parallel}\mathcal{J}_zB_z \right]\\ &&-2\mu_B \left[Q^{\text{eff}}_{\perp}(\mathcal{J}_x^3B_x+\mathcal{J}_y^3B_y)+Q^{\text{eff}}_{\parallel}\mathcal{J}_z^3B_z +  
	\tilde{Q}  \left( B_x \left\{  \mathcal{ J}_x, \mathcal{ J}_z^2 \right\}_s  + B_y \left\{ \mathcal{ J}_y, \mathcal{ J}_z^2 \right\}_s \right) \right]\:, \nonumber 
\end{eqnarray}
where $\left\{ \mathcal{ A} , \mathcal{ B} \right\}_s = \frac12 \left( \mathcal{ A} \mathcal{ B} +  \mathcal{ B}\mathcal{ A} \right)$.	

Note that in Ref. \cite{Semina2023} the last term in Eq. \eqref{effect_Z_D4h} with the parameter $\tilde{Q}$ was omitted. In weak magnetic fields, when the splitting of the heavy ($|M|=3/2$) and light ($|M|=1/2$) holes in zero field is larger than the Zeeman splitting, instead of the parameters $g_{\perp}$, $g_{\parallel}$, $Q^{\text{eff}}_{\perp}$, $Q^{\text{eff}}_{\parallel}$ and $\tilde{Q}$ it is convenient to define the other four parameters
\begin{equation}\label{gfactors_cube_2}
	g_{3/2}^{||}=g_{\parallel}+\frac{9}{2}Q^{\text{eff}}_{\parallel},\quad g_{1/2}^{||}=g_{\parallel}+\frac{1}{2}Q^{\text{eff}}_{\parallel},\quad 
	g_{3/2}^{\perp}=Q^{\text{eff}}_{\perp},\quad g_{1/2}^{\perp}=2 (g_{\perp}+\tilde{Q})+10Q^{\text{eff}}_{\perp},
\end{equation}
having the meaning of longitudinal and transverse $g$ factors of heavy and light holes. A similar situation is realized in quantum wells or quantum wires. In this case, if other effects are not taken into account, the Zeeman splittings of both heavy and light holes will be axially symmetric in the plane of the structure and depend only on the angle $\theta$ between the direction of the magnetic field and the  $z$ axis of the structure \cite{Semina2023}:
\begin{equation}\label{Zeeman_simple}
	|\Delta E_{M}|=2|M|\mu_B\sqrt{(g_{|M|}^{||}B_z)^2+(g_{|M|}^{\perp}B_{\perp})^2}\, ,
\end{equation} 
where $B_{\perp}=\sqrt{B_x^2+B_y^2} = B \sin \theta$. One can see from Eq.~\eqref{gfactors_cube_2} that the contributions of the parameters $g_\perp$ and $\tilde{Q}$ are indistinguishable. Moreover, the transverse $g$ factor of a heavy hole in a structure with the strong heavy-light hole splitting is non-zero only due to the cubic contribution $Q^{\text{eff}}_{\perp}$, which allows one to determine the latter experimentally or by performing numerical calculations. For example, in Ref. \cite{Trifonov} the value of $Q^{\text{eff}}_{\perp}$ was experimentally measured in cylindrical InGaAs quantum dots and its origin due to  the valence band warping  was demonstrated. With a further reduction in symmetry, e.g., to the point group $C_{2v}$, additional contributions may appear, for example, from the interface mixing of heavy and light holes \cite{Ivchenko1996}.

In quantum well structures, the states of heavy and light holes are strongly split in a zero magnetic field. In this case, the heavy-hole $g$ factor is normalized by the admixture of light-hole states due to  the finite width of the quantum well. The results of consistent calculations within the eight-band Hamiltonian, taking into account both the conduction band and the entire valence band, for finite-thickness quantum wells and for superlattices based on A$_3$B$_5$ and A$_2$B$_6$ semiconductors are presented in Ref. \cite{holeKiselev}. As noted above, in quantum wells, the Zeeman splitting of hole states, unlike in spherically symmetric structures, is substantially anisotropic. For example, when the magnetic field is directed in the plane of the interfaces of a narrow quantum well, in which the mixing of heavy and light holes in a zero field can be neglected, the Zeeman splitting of a heavy hole is not equal to zero only due to the anisotropic part of the $g$ factor tensor.

	In thin quantum wells grown along the $z$ axis, the difference between the $g_{zz} \equiv g_{3/2}^\parallel$ component of the $g $ factor tensor of a two-dimensional heavy hole and the $g$ factor of a bulk hole $2\varkappa$ can be calculated in the second order of perturbation theory. The relevant perturbation is the operator proportional to the $z$ component of the momentum operator $\propto \hat{p}_z$ \cite{Wimbauer}, which corresponds to magnetically induced mixing of the ground state of the heavy hole with the excited states of the light hole. In quasi-two-dimensional nanoplatelets or highly oblate quantum dots, the in-plane anisotropy of the structure  has a significant effect on the mixing of states and, thus, on the hole $g$ factor \cite{Semina2021,Semina2015}. The effect of mixing of heavy and light hole states on the hole $g$ factor was experimentally demonstrated in Ref. \cite{Glazman}. It was shown there  that the application of an external electric field that affects the shape of the hole wave function to a structure with SiGe nanocrystals results in a significant modulation of the hole $g$ factor. The effect of the hole state mixing on the components $g_{xx}=g_{yy}=g_{1/2}^\perp$ of the light hole $g$ factor tensor in quantum wells was studied in Ref. \cite{Kiselev2001}.

In calculations, the dependence of the hole  ground state $g$ factor  on the structure parameters turns out to be smooth. However, the same cannot be said about the $g$ factor of excited states. This is related to the mixing of a particular excited state with other states and possible resonances induced by changing the structure parameters. This effect was studied in Refs. \cite{Durnev2012} and \cite{Durnev2014} for the light hole $g$ factor in quantum wells. It was shown that in this case the $g$ factor can reach very large values as the ground subband of a light hole $lh1$ approaches the excited subband of a heavy hole $hh2$. The experimental data presented in Ref. \cite{Durnev2014} confirm this theoretical prediction. Moreover, for excited states, the role of the band structure parameterization and, for example, the effects of hole state mixing at interfaces \cite{Ivchenko1996}, arising from symmetry reduction, significantly increases.

\section{Conclusion} The Land\'{e} factor, or $g$ factor of electrons, holes, and excitons, is a key fundamental parameter in  spin physics. This review successively examines various experimental methods for measuring the electron $g$ factor in bulk semiconductors and semiconductor nanostructures. The following sections present methods for calculating the electron and hole $g$ factors and relate them to the band structure parameters of semiconductors. Methods for studying the Zeeman effect, developed for semiconductors with diamond-like, zinc-blende, and wurtzite structures, are actively applied to investigations of new materials, such as semiconductor perovskites, transition metal dichalcogenides, etc. Analysis of experimental data based on developed theoretical models helps to correctly determine the $g$ factor, for example, in inhomogeneous ensembles of localized excitons \cite{Kotova, Qiang2021, Shornikova2020}. At the same time, a number of experimental results, such as the size dependencies of the electron g factor in CdSe \cite{Zhukov2022} and CuCl \cite{Zhukov2025} NCs or of the hole $g$ factor in perovskite NCs \cite{Nestoklon2023,pernano2025}, cannot be explained by currently developed theoretical models. This may stimulate the search for new unaccounted physical effects hidden behind the observed discrepancies. 

{\bf Acknowledgments}

The authors are grateful to D.R. Yakovlev for careful reading of the manuscript and useful comments. \vspace{2 mm}

{\bf Financial support}

The work of A.V.R. and M.A.S. on the theory of the $g$ factor in nanocrystals was supported by the Russian Science Foundation under project No. 23-12-00300. The work of E.L.I. on the methods of measurements and theory of the $g$ factor was supported by the Russian Science Foundation under project No. 23-12-00142.\vspace{2 mm}

{\bf Conflict of interest}

The authors declare that they have no conflict of interest.

\end{document}